\documentclass[review]{elsarticle}
\usepackage{graphicx}
\usepackage{subfig}
\usepackage{lineno,hyperref}
\usepackage{amsmath, amssymb,amsfonts, mathtools, bm} 
\usepackage{lineno,hyperref}
\usepackage{float}

\journal{Journal of \LaTeX\ Templates}

\begin{document}

\begin{frontmatter}

\title{Ticks on the run: A mathematical model of Crimean-Congo Haemorrhagic Fever (CCHF) -- key factors for transmission\tnoteref{mytitlenote}}
 \tnoteref{mytitlenote}

\author[2]{Suman Bhowmick}
\author[1]{Khushal Khan Kasi}
\author[1]{J\"orn Gethmann}
\author[3]{Susanne Fischer}
\author[1]{Franz J. Conraths}
\author[2,4]{Igor M. Sokolov}
\author[1]{Hartmut H. K. Lentz\fnref{myfootnote}}
\address[1]{Friedrich-Loeffler-Institut, Institute of Epidemiology, S\"udufer 10, 17493 Greifswald, Germany}
\address[2]{Institute for Physics, Humboldt-University of Berlin, Newtonstra{\ss}e 15, 12489 Berlin, Germany}
\address[3]{Friedrich-Loeffler-Institut, Institute of Infectology, S\"udufer 10, 17493 Greifswald, Germany}
\address[4]{IRIS Adlershof, Zum Gro{\ss}en Windkanal 6, 12489 Berlin, Germany}

\fntext[myfootnote]{Corresponding author}
%
%
%

\begin{abstract}
Crimean-Congo haemorrhagic fever (CCHF) is a tick-borne zoonotic disease caused by the Crimean-Congo hemorrhagic fever virus (CCHFV). 
Ticks belonging to the genus \textit{Hyalomma} are the main vectors and reservoir for the virus. 
It is maintained in nature in an endemic vertebrate-tick-vertebrate cycle. 
CCHFV is prevalent in wide geographical areas including  Asia, Africa, South-Eastern Europe and the Middle East.
Over the last decade, several outbreaks of CCHFV have been observed in Europe, mainly in Mediterranean countries.
Due to the  high case/fatality ratio of CCHFV in human sometimes, it is of great importance for public health. 
Climate change and the invasion of CCHFV vectors in Central Europe suggest that the establishment of the transmission in Central Europe may be possible in future.
\par
We developed a compartment-based nonlinear Ordinary Differential Equation (ODE) system to model the disease transmission cycle including blood sucking ticks, livestock and human.
Sensitivity analysis of the basic reproduction number $R_0$ shows that decreasing in the tick survival time is an efficient method to eradicate the disease. 
The model supports us in understanding the influence of different model parameters on the spread of CCHFV.
Tick to tick transmission through co-feeding and the CCHFV circulation through trasstadial and transovarial  stages are important factors to sustain the disease cycle.
The proposed model dynamics are calibrated through an empirical multi-country analysis and multidimensional scaling reveals the disease-parameter sets of different countries burdened with CCHF are different.
This necessary information may help us to select most efficient control strategies.
\end{abstract}


\end{frontmatter}


\section{Introduction}
Crimean-Congo haemorrhagic fever (CCHF) is a tick-borne viral zoonotic disease widely distributed in Asia, Africa, Southeast Europe and the Middle East \cite{10.1093/trstmh/trv050,GARGILI201793}.
CCHF was first identified in an outbreak during World War II on the Crimean Peninsula in 1944--1945, when 200 Soviet military personnel got hemorrhagic fever with a case/fatality ratio of  $10\%$ \cite{Chumakov1945}.
The virus is antigenically identical to a virus that was isolated from the blood of a patient in Democratic Republic of the Congo in 1956. 
The association of these two places resulted in the name of the disease and the virus \cite{Woodall1965congo,CHINIKAR2010110}.
\par
The etiological agent responsible for the disease, i.e. Crimean-Congo hemorrhagic fever virus (CCHFV), belongs to the genus \emph{Orthonairovirus} in the family \emph{Nairoviridae}~\cite{ICTV}.
\par
CCHFV is transmitted between vertebrates and ticks but can also be transmitted horizontally and vertically within the tick population~\cite{10.1371/journal.pntd.0006248, OIE}.
Ticks may be born infected as some fraction of infected female ticks transmit the disease to their offspring.  
Depending on the developmental stage of the ticks, i.e. larvae, nymphs and adults, the vertebrate hosts range from small (birds, hares, rabbits) to large vertebrates (cattle, sheep, humans). 
Animals act as viral amplifying hosts with transient viremia, but they do not develop clinical signs~\cite{GARGILI201793}. 
CCHFV persists in the tick for its whole lifespan~\cite{GARGILI201793}.
Within the tick population, it is transstadially transmitted, venereal transmission among ticks and transmission through co-feeding may also occur~\cite{astmh:/content/journals/10.4269/ajtmh.1989.40.207, GONZALEZ199223}.
The route of transmission through which an infected tick transfer tick-borne pathogens to a susceptible host and vice versa, is described as systemic transmission and another transmission pathway that helps
the pathogens to persist in the ticks through co-feeding, is termed as non-systemic transmission.
\par
Hard ticks of the genus \textit{Hyalomma} are considered the main reservoir and vector for CCHFV~(\cite{astmh:/content/journals/10.4269/ajtmh.1989.40.207, GONZALEZ199223, 10.1371/journal.pntd.0006248}).
They are mainly present in the Afrotropical and in parts of the Palaearctic regions~\cite{Chitimia-Dobler2019,efsa2010scientific,gargili2013influence, ECDC}.
CCHFV has been also detected in other tick genera, including \emph{Rhipicephalus, Amblyomma, Ixodes} and \emph{Dermacentor}, but their role in CCHFV maintenance and their vector capacity is not yet clear~\cite{GARGILI201793}.
\par
CCHF virus is transmitted to people either by tick bites, contact with blood of infected animals or humans, body fluids or tissues. 
Hence, persons involved in the livestock industry, such as agricultural workers, slaughterhouse workers and veterinarians are more vulnerable to CCHFV.
Human-to-human transmission can also happen through contact with bodily fluids of patients comprising virus during the first 7–10 days of illness~\cite{BENTE2013159}. 
Health workers including physicians, nurses and laboratory personnel are therefore at an increased risk of contracting CCHF vectors.
There are several reported cases of nosocomial spread of the disease~\cite{astmh:/content/journals/10.4269/ajtmh.2012.12-0337, Yadav2016, Psh, Nich, Gu} and possibly through sexual transmission~\cite{Nat, OnderErgonul2014}.
Infections acquired in hospitals can also happen due to improper sterilisation of medical equipment or re-use of needles. 
CCHFV represents a potential risk for humans who have unprotected contact with other body fluids~\cite{Bodur}.
Because of severe illness and a high case fatality rate in humans, CCHF is considered as an important vector-borne disease in humans~\cite{WHO1}.
CCHF causes sporadic cases or outbreaks of severe scale across a huge geographical area extending from China to the Middle East, Southeastern Europe and Africa \cite{Zhang, CHINIKAR2010110, 10.1093/trstmh/trv050}. 
It is a highly infectious disease in human with a case fatality rate from $5\%$ to $80\%$~\cite{SAS201838}. 
Human cases are seasonal and associated with an increased population of \textit{H. marginatum} under optimal weather conditions and habitat fragmentation~\cite{estrada}.

According to \cite{BENTE2013159}, the antibody positivity of CCHFV in livestock correlates with the manifestation of human cases and the occurrence of CCHF can happen due to contact with the blood of infected animals.
The authors in \cite{BENTE2013159} mention that repeated outbreaks and sporadic cases have been found in persons handling or slaughtering livestock.
The study conducted in \cite{PMID:5453910}  has shown the infection of a single \textit{H. m. rufipes} that took a blood meal from a viraemic calf.
The authors in \cite{Shephard} demonstrated that adult ixodid ticks of several species can get CCHF virus infection by feeding on viraemic cattle.
In the same study, it was shown that cattle can act as amplifying hosts of CCHFV by viraemic transmission of CCHFV to ticks.
\cite{Shephard} assert that the adult ticks, which get the infection after feeding, conceivably are a crucial source of infection in humans, if pull out by hands or crushed.
The review in \cite{Spen} features the role of livestock in the maintenance and transmission of CCHFV.
The conducted studies in \cite{Atif2017} highlight the fact that the transmission of CCHFV can occur to those who are involved in livestock and animal husbandry in Pakistan, as well as the risk of transmission is higher during the time of Eid-ul-Azha, when Muslims slaughter animals.
According to \cite{Martens}, high prevalence values in ruminants in Turkey depicts the role of these species to identify the high-risk areas.
Additionally, according to the sensitivity analysis of $R_0$ performed in \cite{10.1093/jme/tjy035}, the input of total host density acquires $28\%$ of the total variability and the contribution of hare density is $16\%$ and for the same for the cattle population is $12\%$.\par

\begin{figure}[H]
\centering
\subfloat[Subfigure 1 list of figures text][]{
\includegraphics[width=0.6\textwidth]{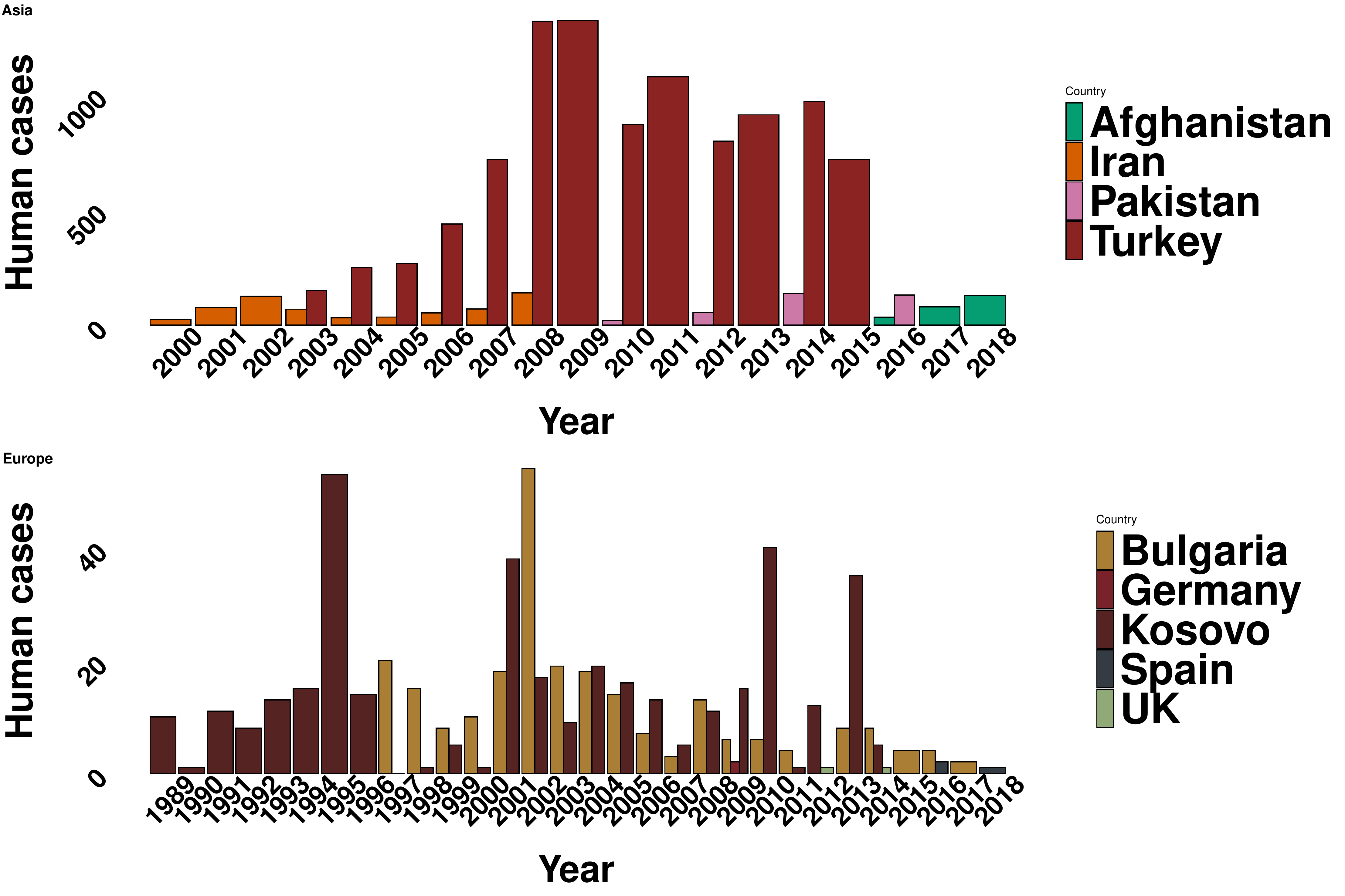}
    \label{Fig:Intro1}
}
\subfloat[Subfigure 2 list of figures text][]{
\includegraphics[width=0.5\textwidth]{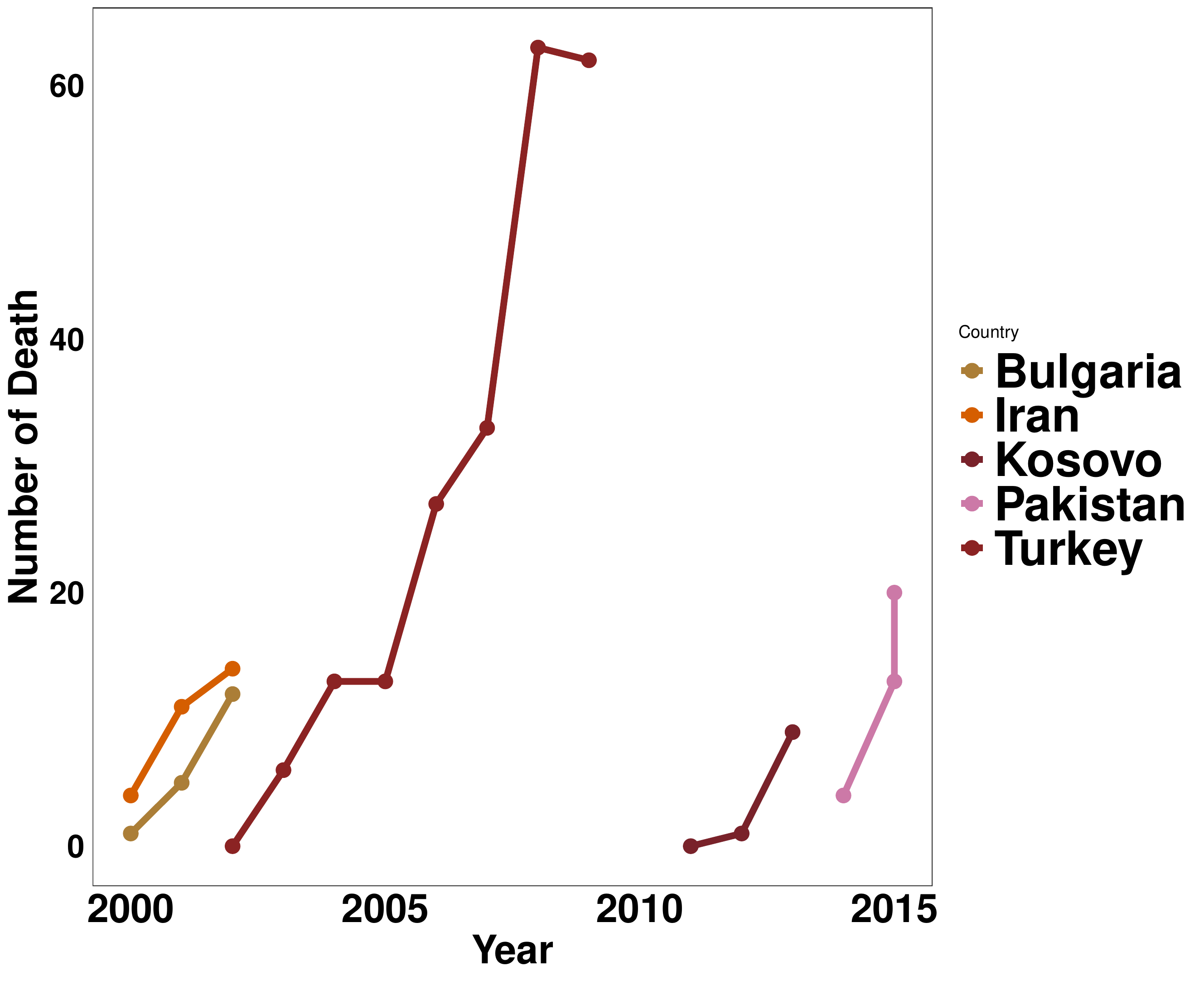}
 \label{Fig:Intro2}
}
\caption{(a) Reported cases of CCHFV in Asia and Europe.
(b) Reported death cases of CCHFV. Data from~\cite{ECDC,10.1371/journal.pone.0110982, Atif2017,WHO}.
 }
\label{Fig:Intro}
\end{figure}

Figure \ref{Fig:Intro} shows reported cases and deaths in humans over time for different countries.
Although the number of cases appears to have stabilised or even decreased in recent years, an increasing trend of fatal cases is shown in Figure \ref{Fig:Intro2}.

\par
CCHFV can spread over long distances through transportation of vectors attached to migratory birds that fly through endemic areas such as Turkey or Greece~\cite{Papa, PASCUCCI2019101272}, or through imported livestock~\cite{Jameson154}.
It has been estimated that every year hundreds of thousands of immature \textit{Hyalomma} ticks are transported via migratory birds into or over Central Europe during the spring migration of birds from southern Europe and Africa~\cite{Chitimia-Dobler2019}.
The virus has a wide range of hosts and vectors and therefore the potential to establish in a new region, if enough susceptible hosts and vectors are available~\cite{doi:10.1111/j.1749-6632.1992.tb19652.x}.
There are several factors like climate change, social and anthropogenic factors that may have contributed to the spread of CCHFV into new regions and to the increase of reported cases~\cite{Expansion}.
Since several years, adult stages \textit{Hyalomma marginatum} ticks have occasionally been found in Germany~\cite{Kampen2007,oehme2017hyalomma,Chitimia-Dobler2019}. 
They may have been introduced by birds as nymphs and continued to develop to the adult stage~\cite{Chitimia-Dobler2019}.
The authors of this study~\cite{Chitimia-Dobler2019} point out  that there is a lack of information about the transportation of \textit{H. marginatum} into Germany and how the tick succeeded to develop into the adult stage in this country. 
In September 2018, successful moulting of a \textit{Hyalomma} nymph removed from a horse in Dorset, England, was reported. 
This horse had no history of overseas travel~\cite{HANSFORD2019704}.
The environmental suitability of CCHFV across Southern and Central Europe has been postulated~\cite{OKELY2019105319}. 
The life cycle of CCHFV is complex and the influence of climate and environment factors has not been fully understand so far.

Mathematical modelling provides a tool to test different scenarios and to analyse the factors influencing the spread of CCHF.
While the number of mathematical models for vector-borne diseases has generally increased, the number of mathematical models for CCHFV in particular is still limited~\cite{doi:10.1080/17513758.2015.1102976, Cooper2007, Hoch2016,10.1093/jme/tjy035}. 
We refine existing models by including human infection.
Humans are an integral part of the transmission cycle and they play a major role in detecting the disease due to the high case fatality ratio in humans.
This will help us to fit the model.
Our current modelling efforts are aiming to provide answers to the following questions: 
\begin{enumerate}
\item  What are the sensitive parameters responsible for the CCHF transmission?
\item What are possible control measures to curb the infection spread in different geographical areas?
\item What is the critical density of ticks necessary for a potential spread
\item How to quantify the nature of dissemination of CCHFV in the endemic areas?
\end{enumerate}
Beside the above mentioned objectives, we utilise the basic reproduction number as a measure of CCHFV transmission potential within the enzootic cycle and this exhibits a crucial feature of risk of human infection.

\section{Model Formulation}
\begin{figure}[H]
    \centering
    \includegraphics[width=0.5\textwidth]{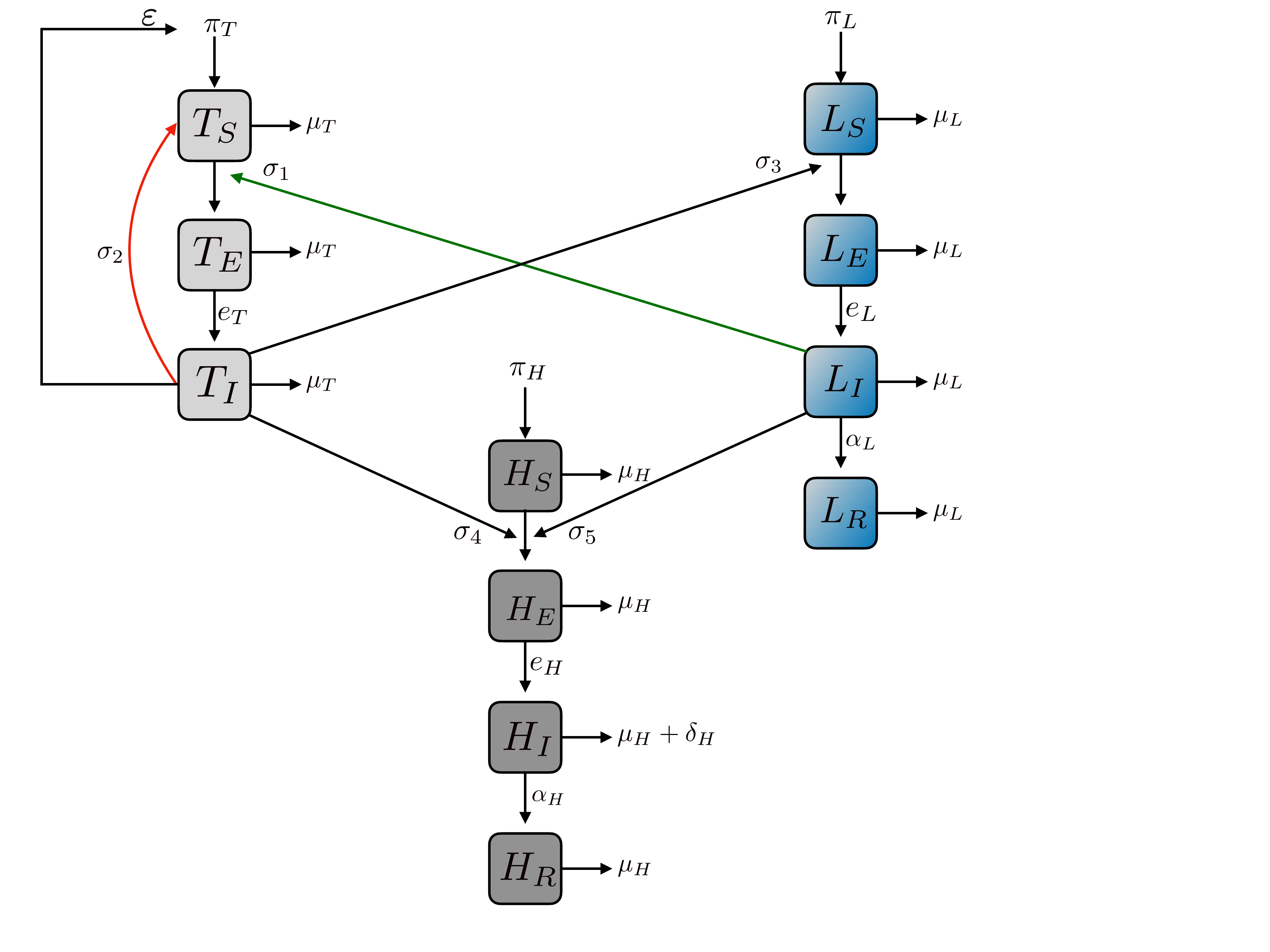}   
    \caption{Infection process among the effective tick population, livestock, and human. 
    The infection process involves non-systemic transmission through co-feeding (red arrow),
    systemic and transovarial transmission (black arrow), acquisition (green arrow). }
    \label{Fig:1}
\end{figure}

In this section first we describe the modelling assumptions and  we then introduce the important model parameters with their meanings.
We set forth our preference is to keep our compartment based Ordinary Differential Equation (ODE) model of three interacting populations, i.e.  ticks, livestock and humans as shown in Figure~\ref{Fig:1} as simple as possible in order to provide general theoretical results and concurrently to abstain from the issue of hyper-parameterisation while taking account of transstadial and transovarial transmissions.
The coupled infection-population model, presented in Figure~\ref{Fig:1}, concisely presents the following mechanisms: transmission (from tick population to host) and acquisition (from livestock to tick population) of the CCHFV pathogen, transstadial persistence of CCHFV amongst the tick life cycle, transovarial transmission from the female adult ticks to eggs and CCHFV transmission from infected livestock to human population along with the net growth of the interacting populations.
After following \cite{BOLZONI2012373},  we summate together all tick stages to reach the equations for the total tick dynamics and admeasure as effective tick population
We assume that there is a fixed proportion of ticks that mature within each developmental stage and the developmental stages of the vectors would function as a delay in the potential infection spread after following the authors in \cite{ROSA2007216}.

Being parsimonious in nature, in our model we club together the different adult tick activity phases together as it paves the way to postulate an abridged model but flexible enough to incorporate further details in the future for possible explorations.
We conveniently attempt to describe the certain biological mechanism involving transstadial transmission, transovarial transmission.
We also assume that systemic infection occurs at the beginning of the blood meal after following the authors in \cite{BOLZONI2012373}.
To accommodate the systemic and non-systemic infection we pursue the efforts done in \cite{BOLZONI2012373, Hoch2016, 10.1093/jme/tjy035}.
In order to  include the CCHFV transmission through co-feeding, we adhere to the following assumptions:
(i) It is proportional to the  number of infected ticks,
(ii) It is proportional to the  number of susceptible ticks,
(iii) Constant number of rodents or the small mammals are being bitten for the blood meals.

Additionally, we make the following assumptions to model the dynamics of CCHFV infection among ticks, livestock, and human:
(i) Homogenous mixing among all the interacting populations at all stages and that CCHFV infection does not alter their movements,
(ii) Livestock has more contact with adult ticks than with other life-stages,
(iii) Infection of \textit{Hyalomma} ticks does not affect the birth or death rates of these ticks,
(iv) Livestock will not die of CCHFV infection~\cite{GARGILI201793} while CCHF-induced deaths in humans are taken into account~\cite{SAS201838}.  

So we en route to construct a mechanistic ODE model including the above knowledge of CCHFV transmission in the human population and to analyse the burden of primary transmission routes undertaking a significant role of dissemination of CCHFV primarily on the effective tick population to simplify the modelling effort.
For the effective tick population, we consider an SEI dynamics as the tick remains infected for life~\cite{GARGILI201793}, while for the livestock and human populations we take an SEIR type of dynamics in to account.

Considering the mentioned assumptions, we end up with the following model system:
For the tick population we have
\begin{eqnarray}\label{eq:1}
           \frac{dT_S}{dt} &=&  \pi_T-\frac{\sigma_1 T_SL_I}{L} -\frac{\sigma_2 T_ST_I}{T}-\mu_T T_S+(1-\varepsilon)\pi_T T_S \\ \nonumber
          \frac{dT_E}{dt} &=& \frac{\sigma_1 T_SL_I}{L} +\frac{\sigma_2 T_ST_I}{T}-\mu_T T_E-e_TT_E\\  \nonumber
          \frac{d T_I}{dt} &=&e_TT_E-\mu_T T_I +\varepsilon \pi_T T_I .
\end{eqnarray}

We model the birth rate $\pi_T = \sigma_T\omega_T \exp\left({-\nu_T\frac{T_S+T_E+T_I}{Ro\omega_1+L\omega_2}}\right)$ after following \cite{ROSA2007216, 10.1093/jme/tjy035},
where $\mathit{T_S}$, $\mathit{T_E}$, $\mathit{T_I}$ are susceptible, exposed and infected ticks respectively, $\nu_T$ is the strength of density-dependence in birth rate,
$\omega_1$ and $\omega_2$ are the weightage of contributions of rodent  and livestock populations on the growth of ticks, $\sigma_T$ is detachment rate of tick and $\omega_T$ is mean no of eggs laid by an adult female tick, $Ro$ is the constant number of total rodents and $L$ is the total number of livestock.  
We incorporate the rodent-tick transmission cycle without explicitly deriving the equations of different stages of ticks.
In order to include the transovarial transmission from adult female ticks to eggs, we introduce a parameter $\varepsilon$ that measures the proportion of infected eggs laid by an infected female adult tick as mentioned in \cite{ROSA2007216}.
We have also opted for a density-dependent mortality rate ($\mu_T$) according to \cite{10.1093/jme/tjy035}.
The transmission parameters of tick related model system are as follows:
Livestock to tick infection transmission rate has been modelled according to \cite{doi:10.1111/j.1461-0248.2009.01378.x, Hoch2016} as 
$\sigma_1 = \frac{p_T\gamma_TN_Tf_S}{d_T}$, where $p_T$ is defined as transmission efficiency from livestock to tick, $\gamma_T$ is the duration of infective period,
$N_T$ is the rate of average no of feeding ticks on livestock, $f_S$ is defined as the fraction of blood meal per tick and $d_T$ is duration of attachment.
Non-systemic transmission term $\sigma_2 $ is defined according to  \cite{BOLZONI2012373} as following:
$\sigma_2 = [1- \exp\left\{  -(n_{Ro}+l_{Ro})\theta \right\}]\sigma_{Ro} Ro$, where
$n_{Ro}$ is the fraction of nymphs against the total number of ticks feeding on the rodents, $l_{Ro}$ is the fraction of larvae against the total number of ticks feeding on the rodents, 
$\theta$ is the transmission probability, $\sigma_{Ro}$ is encounter rate between the ticks and rodents.
The domestic livestock population is described by the following system of equations:
\begin{eqnarray}\label{eq:2}
    	  \frac{dL_S}{dt} &=&  \pi_L-\frac{\sigma_3 L_ST_I}{L} -\mu_L L_S \\ \nonumber
    	   \frac{dL_E}{dt} &=& \frac{\sigma_3 L_ST_I}{L}-e_LL_E-\mu_L L_E\\ \nonumber
   	   \frac{dL_I}{dt} &= &e_LL_E-\alpha_LL_I-\mu_L L_I \\ \nonumber
    	   \frac{d L_R}{dt} &=&\alpha_LL_I-\mu_L L_R .
\end{eqnarray}
And the human population is described by the following system of equations:
\begin{eqnarray}\label{eq:3}
    	  \frac{dH_S}{dt}&=&  \pi_H-\frac{\sigma_4 H_ST_I}{H} -\frac{\sigma_5H_SL_I}{H} -\mu_H H_S \\ \nonumber
	  \frac{dH_E}{dt}&=& \frac{\sigma_4 H_ST_I}{H} +\frac{\sigma_5H_SL_I}{H}-e_HH_E-\mu_H H_E\\ \nonumber
	 \frac{dH_I}{dt} &=& e_HH_E-\alpha_HH_I-\mu_H H_I-\delta_H H_I \\ \nonumber
	 \frac{dH_R}{dt} &=& \alpha_HH_I-\mu_H H_R .
\end{eqnarray}

Here $\mathit{L_S}$, $\mathit{L_E}$, $\mathit{L_I}$ and $\mathit{L_R}$ represent susceptible, exposed, infected and recovered livestock and similarly $\mathit{H_S}$, $\mathit{H_E}$, $\mathit{H_I}$ and $\mathit{H_R}$ do the same for  human population.
We also have assumed a density-dependent birth rate ($\pi_L$) for the livestock and the linear growth rate ($\pi_H$) for human.
We model the acquisition rate $\sigma_3  = [1- \exp\left\{- (N_{Ro}\kappa_{N}+L_{Ro}\kappa_{L}+A_{L}\kappa_{A}) \right\}]\sigma_{L}L$ after following \cite{Hoch2016, 10.1093/jme/tjy035} to include the propagation of infection acquired during the transstadial stages.
$\kappa_i$ is the transmission rate from larvae, nymphs and adult ticks, where $i = L, N, A$.
$N_{Ro}$ is the ratio between the infectious nymphs and constant rodents density,
$L_{Ro}$ is the ratio between the infectious larvae and constant rodents density,
$A_{L}$ is the ratio between the infectious adult ticks and livestock density and 
$\kappa_i = 1-(1-T_i)^{\frac{1}{d_{feed_i}}}$, with $T_i$ is the over all efficiency of transmission and $d_{feed_{i}}$ is the feeding duration and $\sigma_{L}$ is encounter rate between the ticks and livestock.
$\sigma_4$ is the transmission rate from an infected tick to a susceptible human and $\sigma_5$ is the transmission rate from an infected livestock to a susceptible human. 
A full list of model parameters, variables and their biological meanings are given in Tables \ref{table:1}, \ref{table:2} and \ref{table:3} as well as in Supplementary Information (SI).

\begin{table}[H]
\centering
\begin{tabular}{||c c ||} 
 \hline
 \tiny{Variable} & \tiny{Description of Model Variables} \eqref{eq:1}, \eqref{eq:2}, \eqref{eq:3},  \eqref{eq:H3b}  \\ [0.5ex] 
 \hline\hline
   \tiny{$\mathit{T_S}$}&  \tiny{Susceptible  ticks}  \\ 
  \tiny{$\mathit{T_E}$}  &  \tiny{Exposed  ticks}   \\
  \tiny{$\mathit{T_I}$} & \tiny{ Infected  ticks}   \\
  \tiny{$\mathit{L_S}$}  &  \tiny{Susceptible livestocks}   \\
  \tiny{$\mathit{L_E}$}  &  \tiny{Exposed livestocks}   \\  
  \tiny{$\mathit{L_I}$}  &  \tiny{Infected  livestocks} \\ 
  \tiny{$\mathit{L_R}$}  &  \tiny{Recovered livestocks}   \\ 
  \tiny{$\mathit{H_S}$}  &  \tiny{Susceptible humans}   \\
  \tiny{$\mathit{H_E}$}  &  \tiny{Exposed  humans}  \\ 
  \tiny{$\mathit{H_I}$}  &  \tiny{Infected  humans} \\ 
  \tiny{$\mathit{H_R}$}  &  \tiny{Recovered humans}  \\ 
  \tiny{$\mathit{T}$ } &  \tiny{Total tick population}  \\ 
  \tiny{$\mathit{L}$}  &  \tiny{Total livestock population}   \\ 
  \tiny{$\mathit{Ro}$ } &  \tiny{Total rodent population}   \\ 
  \tiny{$\mathit{H}$}  &  \tiny{Total human population}   \\ 
  \tiny{$\sigma_1$}  &  \tiny{Transmission parameter: livestock to tick}  \\ 
  \tiny{$\sigma_2$}  &  \tiny{Transmission parameter: tick to tick}  \\ 
  \tiny{$\sigma_3$}  &  \tiny{Transmission parameter: tick to livestock}  \\ 
  \tiny{$\sigma_4$}  &  \tiny{Transmission parameter: tick to human}  \\  
  \tiny{$\sigma_5$}  &  \tiny{Transmission parameter: livestock to human}  \\
 \tiny{$\sigma_6$} &  \tiny{Transmission parameter: human to human}  \\ 
  \tiny{$\varepsilon$} &  \tiny{proportion of the newborn infected ticks}  \\ [1ex]
 \hline
\end{tabular}
\caption{Variables used in the model \eqref{eq:1}, \eqref{eq:2}, \eqref{eq:3},  \eqref{eq:H3b}.}
\label{table:1}
\end{table}

\begin{table}[H]
\centering
\begin{tabular}{c c c  c} 
 \hline
  \tiny{Parameter} & \tiny{ Description}  &  \tiny{Range}  &  \tiny{References}\\ [0.5ex] 
 \hline\hline 
 \tiny{ $\pi_L$}  & \tiny{Birth term of livestock population}&  \tiny{$\left[0.5,  1.5\right]$}    &~\tiny{\cite{{Mpeshe2011}}} \\ [1ex]
  \tiny{$\pi_H$}  & \tiny{Birth term of human population} &  \tiny{$\left[0.5,  1.5\right]$}    &~\tiny{\cite{Mpeshe2011}} \\ [1ex]
  \tiny{$\mu_L$}  & \tiny{Death term of livestock population} &  \tiny{$\left[1/3600,  1/360\right]$}    &~\tiny{\cite{Mpeshe2011}} \\ [1ex]
  \tiny{$\mu_H$}  & \tiny{Death term of human population} &  \tiny{$\left[1/365\times 60, 1/365\times 40 \right]$}  & ~\tiny{\cite{Mpeshe2011}} \\ [1ex]
 \tiny{ $1/e_T$}  & \tiny{Incubation period in  tick}&  \tiny{$\left[1,  3\right]$}   &~\tiny{\cite{doi:10.1111/j.1461-0248.2009.01378.x,10.1309/LMN1P2FRZ7BKZSCO}} \\ [1ex]
  \tiny{$1/e_L$}  & \tiny{Incubation period in  livestock} &  \tiny{$\left[3,  5\right]$}   &~\tiny{\cite{doi:10.1111/j.1461-0248.2009.01378.x}} \\ [1ex]
  \tiny{$1/e_H$}  & \tiny{Incubation period in  human} &  \tiny{$\left[1,  9\right]$}   &~\tiny{\cite{10.1309/LMN1P2FRZ7BKZSCO}} \\ [1ex]
  \tiny{$\sigma_2$}  & \tiny{Transmission parameter: tick to tick} &  \tiny{$\left[0.01,  0.04\right]$}    &~\tiny{\cite{doi:10.1111/j.1461-0248.2009.01378.x}} \\ [1ex]
  \tiny{$\sigma_4$}  & \tiny{Transmission parameter: tick to human} &  \tiny{$\left[0.25, 0.375\right]$}  & \tiny{\cite{Kriesel2009MathematicalMO}} \\ [1ex]
  \tiny{ $\sigma_5$}  & \tiny{Transmission parameter: livestock to human} &  \tiny{$\left[0.001, 0.002\right]$}    &~\tiny{\cite{Mpeshe2011}} \\ [1ex]
  \tiny{$\beta_H$}  & \tiny{Effective contact rate: human to human} &  \tiny{$\left[0.5,  0.75\right]$}   &~\tiny{\cite{doi:10.1080/02286203.2017.1320820}} \\ [1ex]
  \tiny{$\eta_H$}  & \tiny{Proportion of quarantined} &  \tiny{$\left[0.0005,  0.0075\right]$}   &~\tiny{\cite{doi:10.1080/02286203.2017.1320820}} \\ [1ex]
  \tiny{$\lambda_H$}  & \tiny{Surveillance coverage} &  \tiny{$0.85$}   &~\tiny{\cite{doi:10.1080/02286203.2017.1320820}} \\ [1ex]
  \tiny{$\tau_H$}  & \tiny{Availability of isolation centres} &  \tiny{$0.65$}   &~\tiny{\cite{doi:10.1080/02286203.2017.1320820}} \\ [1ex]
  \tiny{$\gamma_H$}  & \tiny{Enhanced personal hygiene} &  \tiny{$\left[0,  0.075\right]$}   &~\tiny{\cite{doi:10.1080/02286203.2017.1320820}} \\ [1ex]
  \tiny{$\sigma_H$}  & \tiny{Rate of public enlightenment} &  \tiny{$0.90$}   &~\tiny{\cite{doi:10.1080/02286203.2017.1320820}} \\ [1ex]
 \tiny{ $1/\alpha_L$}  & \tiny{Recovery period of livestock} &  \tiny{$\left[14,  21\right]$}   &~\tiny{\cite{10.1093/jme/tjy035}} \\ [1ex]
 \tiny{ $1/\alpha_H$}  & \tiny{Recovery period of human population} &  \tiny{$\left[15,  21\right]$}   &~\tiny{\cite{Papa2002, 10.1309/LMN1P2FRZ7BKZSCO}} \\ [1ex]
  \tiny{$\delta_H$}  & \tiny{Disease induced death } &  \tiny{$\left[0.3,  0.8\right]$}    &~\tiny{\cite{Schuster2016, 10.1309/LMN1P2FRZ7BKZSCO}} \\ [1ex]
    \hline
\end{tabular}
\caption{Variables used in the model \eqref{eq:1}, \eqref{eq:2}, \eqref{eq:3}, \eqref{eq:3a},  \eqref{eq:H3b}.}
\label{table:2}
\end{table}
\begin{table}[H]
\centering
\begin{tabular}{c c c  c} 
 \hline
  \tiny{Parameter} & \tiny{ Description}  &  \tiny{Range}  &  \tiny{References}\\ [0.5ex] 
 \hline\hline 
\tiny{$\omega_T$}  & \tiny{Mean no of eggs} &   \tiny{$\left[4258, 9476\right]$}   & \tiny{\cite{doi:10.1111/j.1461-0248.2009.01378.x}}  \\ [1ex]
   \tiny{$\nu_T$}  & \tiny{Strength of density-dependence in birth rate} &   \tiny{$0.025$}    & \tiny{\cite{BOLZONI2012373, ROSA2007216}}  \\ [1ex]
   \tiny{$\sigma_T$}  & \tiny{Detachment rate of tick} &   \tiny{$0.256$}    & \tiny{\cite{doi:10.1111/j.1461-0248.2009.01378.x}}  \\ [1ex]
   \tiny{$\omega_1$}  & \tiny{Contribution of rodent population } &   \tiny{$0.4$}    & \tiny{\cite{ROSA2007216}}  \\ [1ex]
   \tiny{$\omega_2$}  & \tiny{Contribution of livestock population } &   \tiny{$0.04$}   & \tiny{\cite{ROSA2007216}}  \\ [1ex]
   \tiny{$p_T$}  & \tiny{Transmission efficiency: livestock to tick} &   \tiny{$\left[0.11, 0.33\right]$}   & \tiny{\cite{doi:10.1111/j.1461-0248.2009.01378.x}}  \\ [1ex]
   \tiny{$\gamma_T$}  & \tiny{Duration of infective period} &   \tiny{$\left[2, 6\right]$}   & \tiny{\cite{doi:10.1111/j.1461-0248.2009.01378.x}}  \\ [1ex]
   \tiny{$d_T$}  & \tiny{Duration of attachment} &   \tiny{$\left[6, 8\right]$}   & \tiny{\cite{doi:10.1111/j.1461-0248.2009.01378.x}}  \\ [1ex]
   \tiny{$N_T$}  & \tiny{Rate of average no of feeding ticks on livestock} &   \tiny{$\left[5.5, 8.5\right]$}   & \tiny{\cite{doi:10.1111/j.1461-0248.2009.01378.x}}  \\ [1ex]
 \tiny{$\mu_0$} & \tiny{Basal mortality rate} &   \tiny{$0.28$}   & \tiny{\cite{10.1093/jme/tjy035}}  \\ [1ex]
  \tiny{$\alpha_0$} & \tiny{Influence of density-dependence on mortality } &   \tiny{$0.1$}   & \tiny{\cite{10.1093/jme/tjy035}}  \\ [1ex]
  \hline
\end{tabular}
\caption{Variables used in the model \eqref{eq:1}, \eqref{eq:2}, \eqref{eq:3}, \eqref{eq:3a}, \eqref{eq:H3b}.}
\label{table:3}
\end{table}

\section{Basic Reproduction Number $R_{0}$ }\label{R0}
In the course of epidemiology, the spread of epidemics often described by an epidemiological metric called as \emph{basic reproduction number} ($R_0$).
Elementarily, it characterises the expected number of secondary cases produced by a single primary case in a completely susceptible population.
To illustrate the spread of pathogens that infect multiple hosts, \cite{Diekmann} incepted a formal mathematical framework named as \emph{Next Generation Matrix} (NGM).
The elements of NGM ($K_{ij}$) are the expected number of infected of type $i$ produced by a single infectious individual of type $j$.
Mathematically speaking, the matrix NGM depicts the linearisation of  the nonlinear model system when all the hosts are of susceptible type.
The largest eigenvalue of the NGM is defined as the basic reproduction number.\par
For the tick-borne disease the definition of $R_0$ is slightly different.
In this case, $R_0$ is described as number of new female parasites produced by a female parasites without considering the density-dependent constraints governing the life cycle of parasites \cite{ROSA2007216}.
First, we address the question, under which conditions the virus can spread in an initially susceptible population, if a single infected individual is introduced. 
Mathematically, we analyse the stability of the disease-free equilibrium $E_0$, which is a fixed point of the system \eqref{eq:1}, \eqref{eq:2}, \eqref{eq:3}.
It is given by: 
$
E_0 = \left(T_S^*, 0, 0, L_S^*, 0, 0, 0, H_S^*, 0, 0, 0\right) =  \left(\frac{\pi_T^0}{\mu_T}, 0, 0, \frac{\pi_L}{\mu_L}, 0, 0, 0, 0, \frac{\pi_H}{\mu_H}, 0, 0, 0, 0\right). 
$

Information regarding $E_0$, the mathematical properties of behaviour of the model solution and the stability analysis are appended in the Supplementary Information (SI).
If the disease-free equilibrium $E_0$ is stable, the disease dies out before it can infect individuals, and it can spread over the population if $E_0$ is instable.
The stability condition for $E_0$ can be expressed in terms of  $R_0$, where the outbreak condition is $R_0>1$ \cite{keeling2011modeling}.
To compute the basic reproduction number, we use the NGM method  as described in \cite{VANDENDRIESSCHE200229}.
Detailed computation is included in the Supplementary Information (SI).

The next generation matrix can be written as following:
\begin{equation} \label{matK}
\mathcal{K} =      	
\begin{pmatrix}
\frac{\mathit{T_S^*} \mathit{e_T} \sigma_{2}}  {T {\left(\mathit{e_T} + \mu_T\right)}  {\left(\mu_T-\varepsilon \pi_T\right)}}  &  
\frac{\mathit{T_S^*} \mathit{e_L} \sigma_{1}}{L {\left(\alpha_L + \mu_L\right)} {\left(\mathit{e_L} + \mu_L\right)}}   \\
\frac{\mathit{L_S^*} \mathit{e_T} \sigma_{3}}{L {\left(\mathit{e_T} + \mu_T\right)}  {\left(\mu_T-\varepsilon \pi_T\right)}}  & 0  \\ 
\end{pmatrix}.
\end{equation}
The matrix $\mathcal{K}$ \eqref{matK} can be biologically interpreted as 
\begin{equation} \label{KInter}  	
 \mathcal{K} = 
\begin{pmatrix}
Tick \hookrightarrow Tick & Livestock \hookrightarrow Tick  \\
Tick \hookrightarrow Livestock & 0  \\
\end{pmatrix} ,
\end{equation}
where $ X \hookrightarrow Y$ means population $X$ is infecting population $Y$.

The basic reproduction number is defined as the spectral radius of the NGM \eqref{matK}.
In our model we decompose the total basic reproduction number $R_0$ into different contributions.
These are (i) infection from tick to tick via co-feeding and vertical transmission ($R_T$) and (ii) infection from tick to livestock model system $R_{LA}$.
For the whole model we get
\begin{eqnarray}\label{R_LA}
R_0 &=& \frac{R_{T}}{2} +\sqrt{\left(\frac{R_{T}}{2}\right)^2 +R_{LA}} \\ 
&\implies&  \frac{1}{2} \left[R_T+\sqrt{R_T^2+4R_{LA}}\right] \nonumber
\end{eqnarray}
where
\begin{equation}\label{RT}
R_{T}=  
\left[ \frac{\pi_T^{0}} {T} \frac{\mathit{e_T}} {\left(\mathit{e_T} + \mu_T\right)} \frac{\sigma_{2}} {\varepsilon \pi_T^{0}-\mu_T}\frac{1}{(1-\varepsilon)\pi_T^{0}-\mu_T}\right]  
\end{equation}
is the contribution of tick-to-tick transmission due to co-feeding and transovarial transmission  and 
\begin{equation}\label{RTL}
\small{R_{LA}=  \left[\left(\frac{\pi_T^0}{L} \frac{\mathit{e_L}} {\left(\mathit{e_L}+ \mu_L\right)} \frac{1} {\left(\alpha_L + \mu_L\right)}\frac{\sigma_{1}} {\mu_L}\right)
 \left( \frac{\pi_L} {L} \frac{\mathit{e_T}} {\left(\mathit{e_T} + \mu_T\right)}  \frac{\sigma_{3}} {\varepsilon \pi_T^0-\mu_T}\frac{1} {(1-\varepsilon)\pi_T^0-\mu_T}\right)   
 \right]}
\end{equation}
is the contribution of tick-to-livestock and livestock-to-tick transmission.
The equation \eqref{R_LA} can also be represented as :
\begin{equation}\label{R_LA1}
 \small{R_0 = \frac{Tick \hookrightarrow Tick}{2} +\sqrt{\left(\frac{Tick \hookrightarrow Tick}{2}\right)^2 +(Livestock \hookrightarrow Tick)(Tick \hookrightarrow Livestock)} },
\end{equation}
If we exclude the transmission through co-feeding then the basic reproduction number is simply
$R_0 ^\mathrm{w} = \sqrt{R_{LA}}$,
where the index $\mathrm{w}$ stands for without co-feeding.
The epidemic threshold is the critical point, where $R_0 = 1$,
$R_0 ^\mathrm{w}$ can biologically be described as  $ R_0 ^\mathrm{w}= \sqrt{\left(Livestock \hookrightarrow Tick)(Tick \hookrightarrow Livestock\right)}$,
It follows from  \eqref{R_LA} that at the critical point the contributions of both transmission ways simply add up, i.e.
$R_{T}^{C}+R_{LA}^{C} =1 $.
The terms in \eqref{RT} can be interpreted as follows: $\frac{\mathit{e_T}} {\mathit{e_T} + \mu_T}$ is the probability that a tick will survive the incubation period and become infectious after co-feeding, 
 $\frac{1} {(1-\varepsilon)\pi_T^0-\mu_T}$ is the natural growth of a susceptible tick, $\frac{\sigma_{2}} {\varepsilon \pi_T^0-\mu_T}$ is the probability of CCHFV transmission from a tick to another tick through non-systemic and transstadial in its lifetime,
 $\frac{\pi_T^0} {T}$ is the ratio between the birth rate of a tick and the total number of ticks.
In the same way the terms in \eqref{RTL} can be explained as before. 
$\frac{\mathit{e_L}} {\mathit{e_L}+ \mu_L}$ is the the proportion of livestock that will survive the incubation period and become infectious,
$ \frac{1} {\alpha_L + \mu_L}$ is the infectious lifespan of this livestock, $\frac{\sigma_{1}} {\mu_L}$ is the probability of CCHFV transmission from the livestock to a tick in the lifespan of infectious livestocks,
$\frac{\sigma_{3}} {\varepsilon \pi_T^0-\mu_T}$ is the probability of CCHFV transmission from an infectious tick to livestock during the span of its natural growth, 
$\frac{\pi_T^0} {L}$ is the ratio between the birth rate of tick 
and the total number of livestock and $\frac{\pi_L} {L}$ is the ratio between the birth rate of livestock and the total number of livestock.
Finally, using the parameter values provided in \ref{table:2}, we obtain the following figures for the basic reproduction number \eqref{R_LA}
\begin{equation}
R_0 = 3.4 ,
\end{equation}
where the contributions are for the co-feeding $R_T = 1.6$, and for the tick-to-livestock and livestock-to-tick infection $R_{LA} = 7.2$.
The chosen parameters are the minima of the respective parameter ranges.
When we perform the same calculations with the assumed maximum values of the parameters, we get $R_T =2.4$, $R_{LA}= 10.75$ and $R_0= 4.9$

\subsection{Inclusion of human-to-human transmission}
\begin{figure}[H]
\centering
  \includegraphics[width=0.6\textwidth]{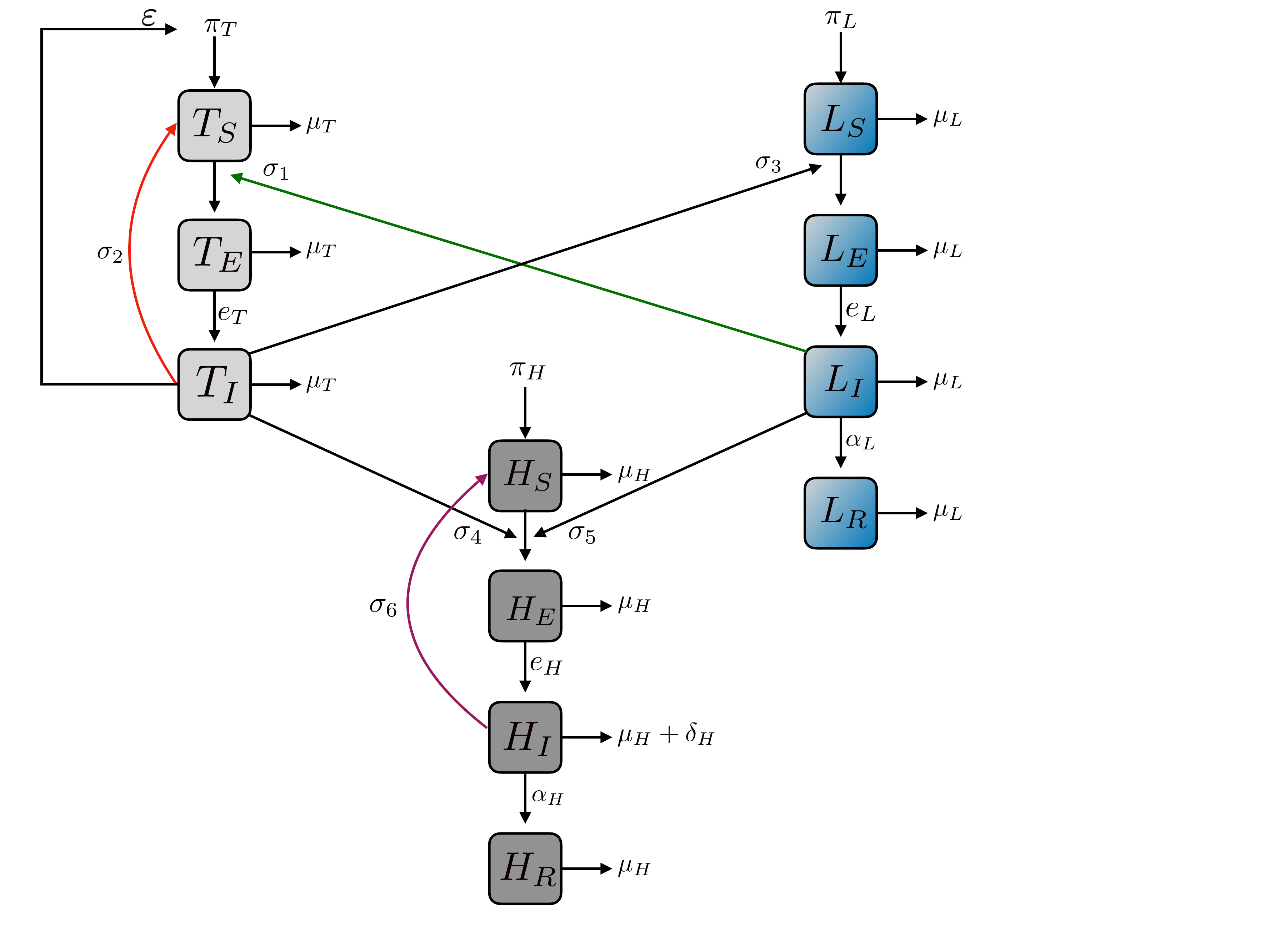}   
\caption{
Flow chart explicating the same infection process as mentioned in Figure~\ref{Fig:1} with the inclusion of human to human transmission (described in maroon colour arrow).
}
\label{Fig:2}
\end{figure}

CCHFV is a viral zoonosis with cases of human-to-human transmission \cite{Onder} and case/fatality ratios ranging between $5\%$ to $80\%$~\cite{SAS201838}.
To take account of the nosocomial spread of CCHFV and cases of human-to-human transmittal, we include another transmission route as human-human transmission ($\sigma_6$) \cite{Garrison} in the model as depicted in Figure~\ref{Fig:2}. 
The description of $\sigma_6$ is included in SI.
So, our model equation system described in \eqref{eq:3} modifies into the following:

\begin{eqnarray}\label{eq:3a}
           \frac{dH_S}{dt}&=&  \pi_H-\frac{\sigma_4 H_ST_I}{H} -\frac{\sigma_5H_SL_I}{H}-\frac{\sigma_6H_SH_I}{H} -\mu_H H_S \\ \nonumber
	 \frac{dH_E}{dt}&=& \frac{\sigma_4 H_ST_I}{H} +\frac{\sigma_5H_SL_I}{H}+\frac{\sigma_6H_SH_I}{H}-e_HH_E-\mu_H H_E\\ \nonumber
	 \frac{dH_I}{dt} &=& e_HH_E-\alpha_HH_I-\mu_H H_I-\delta_H H_I \\ \nonumber
	 \frac{dH_R}{dt} &=&\alpha_HH_I-\mu_H H_R .
\end{eqnarray}

This system has a next generation matrix that can be simplified to
\begin{equation}\label{MatK}   	
 \mathcal{K} = 
\begin{pmatrix}
\frac{\mathit{T_S^*} \mathit{e_T} \sigma_{2}}{T {\left(\mathit{e_T} + \mu_T\right)} \mu_T} & \frac{\mathit{T_S^*} \mathit{e_L} \sigma_{1}}{L {\left(\alpha_L + \mu_L\right)} {\left(\mathit{e_L} + \mu_L\right)}} & 0 \\
\frac{\mathit{L_S^*} \mathit{e_T} \sigma_{3}}{L {\left(\mathit{e_T} + \mu_T\right)} \mu_T} & 0 & 0 \\
\frac{\mathit{H_S^*} \mathit{e_T} \sigma_{4}}{L {\left(\mathit{e_T} + \mu_T\right)} \mu_T} & \frac{\mathit{H_S^*} \mathit{e_L} \sigma_{5}}{H {\left(\alpha_L + \mu_L\right)} {\left(\mathit{e_L} + \mu_L\right)}} & \frac{\mathit{H_S^*} \mathit{e_H} \sigma_{6}}{H {\left(\alpha_H + \delta_H + \mu_H\right)} {\left(\mathit{e_H} + \mu_H\right)}}
\end{pmatrix}
\end{equation}
with spectral radius
\begin{equation}\label{RH,RLA}
R_0 = \max{[R_H,  R_{LA}]}
\end{equation}
where  \, \, 
\begin{equation}\label{RH}
R_H =\left[\frac{\pi_H}{H}\frac{\sigma_{6}}{\mu_H}\frac{\mathit{e_H}}{\left(\mathit{e_H} + \mu_H\right)}\frac{1}{\left(\alpha_H + \delta_H + \mu_H\right)}\right]
\end{equation}

The matrix $\mathcal{K}$ \eqref{MatK} can be biologically interpreted as 

\begin{equation} \label{KInter1}  	
 \mathcal{K} = 
\begin{pmatrix}
Tick \hookrightarrow Tick & Livestock \hookrightarrow Tick & 0 \\
Tick \hookrightarrow Livestock & 0 & 0 \\
Tick \hookrightarrow Human & Livestock \hookrightarrow Human & Human \hookrightarrow Human \\
\end{pmatrix}
\end{equation}

\subsection{Tick-Human Model}\label{Tick-Human}
According to \cite{BENTE2013159, VOROU2009659, doi:10.1080/17513758.2015.1102976}, many of the reported cases of CCHFV are due to the bites by adult ticks.
As reported in \cite{BENTE2013159}, \textit{Hyalomma} are "hunting" ticks and they  can chase up to 400 m to find their hosts (including humans).
A survey conducted in Turkey reveals that among all reported cases, $68.9\%$ had a history of tick-bites or the contact with ticks, while only $0.16\%$ cases represented nosocomial infections \cite{Yilmaz2009}. 
The authors in \cite{Mourya} mention that humans might also be infected due to occupational exposure to bites by infected ticks or crushing infected ticks with bare hands.
The study in \cite{ATKINSON2013e1031} reported a large number of patients who tested positive for CCHF due to potential exposure via tick bites along with asymptomatic cases of CCHF in Tajikistan.
After following the edict of WHO, \cite{WHO3} it is often possible to reduce or to curtail the risk of animal-to-human transmission in the countries with  better health care facilities, available techniques and hygiene practices in slaughtering, meat handling, quality control and the widespread awareness of the perils of CCHFV transmission,  whereas for the countries lacking such resources this can be a daunting task to follow \cite{Aslam}.
In order to mimic CCHF transmission under ideal hygiene conditions (slaughtering and meat handling), we consider only the subsystem related to the humans and the ticks of the model system while ignoring the livestock-to-human transmission path \eqref{eq:1}, \eqref{eq:2}, \eqref{eq:3a}.

We obtain the following ODE system:

\begin{eqnarray}\label{eq:T1a}
          \frac{dT_S}{dt}&=& \pi_T-\frac{\sigma_2 T_ST_I}{T}-\mu_T T_S \\ \nonumber
 	\frac{dT_E}{dt}&=& \frac{\sigma_2 T_ST_I}{T}-\mu_T T_E-e_TT_E\\ \nonumber
 	\frac{d T_I}{dt} &=& e_TT_E-\mu_T T_I .
\end{eqnarray}

\begin{eqnarray}\label{eq:H3b}
          \frac{dH_S}{dt}&=&  \pi_H-\frac{\sigma_4 H_ST_I}{H} -\frac{\sigma_6H_SH_I}{H} -\mu_H H_S \\ \nonumber
	 \frac{dH_E}{dt}&=& \frac{\sigma_4 H_ST_I}{H} +\frac{\sigma_6H_SH_I}{H}-e_HH_E-\mu_H H_E\\ \nonumber
	 \frac{dH_I}{dt} &=& e_HH_E-\alpha_HH_I-\mu_H H_I-\delta_H H_I \\ \nonumber
 	\frac{dH_R}{dt} &=& \alpha_HH_I-\mu_H H_R .
\end{eqnarray}

Next generation matrix $(\mathcal{K_{TH}})$ associated with \eqref{eq:T1a}, \eqref{eq:H3b} is given by
\begin{equation}   	
 \mathcal{K_{TH}} =      	
\begin{pmatrix}
\frac{\mathit{T_S^*} \mathit{e_T} \sigma_{2}}{T {\left(\mathit{e_T} + \mu_T\right)} \mu_T} & \frac{\mathit{T_S^*} \sigma_{2}}{T \mu_T} & 0 & 0 \\
0 & 0 & 0 & 0 \\
\frac{\mathit{H_S^*} \mathit{e_T} \sigma_{4}}{H {\left(\mathit{e_T} + \mu_T\right)} \mu_T} & \frac{\mathit{H_S^*} \sigma_{4}}{H \mu_T} & \frac{\mathit{H_S^*} \mathit{e_H} \sigma_{6}}{H {\left(\alpha_H + \delta_H + \mu_H\right)} {\left(\mathit{e_H} + \mu_H\right)}} & \frac{\mathit{H_S^*} \sigma_{6}}{H {\left(\alpha_H + \delta_H + \mu_H\right)}} \\
0 & 0 & 0 & 0
\end{pmatrix}
\end{equation} 
\begin{equation}\label{RTH}
 R_{TH}= \max{[R_H,  R_{T}]}
\end{equation}
When we include the whole model but exclude the livestock to humans transmission then we can have the basic reproduction number of the decoupled system as $R_{TH}= \max{[R_H,  R_{LA}]}$.
Once again following (\cite{NGUYEN201928}), we replace the host-specific transmission rate ($\sigma_4 \rightarrow  \zeta\sigma_4, \sigma_6 \rightarrow  \zeta\sigma_6 $) and 
calculate the value of $R_{H}$ where $\zeta \in [0, 1]$.
$R_H$ becomes less than one when $\zeta \approx .0.37$
\section{Dynamics of the model}
\begin{figure}[H]
    \centering
    \includegraphics[width=0.6\textwidth]{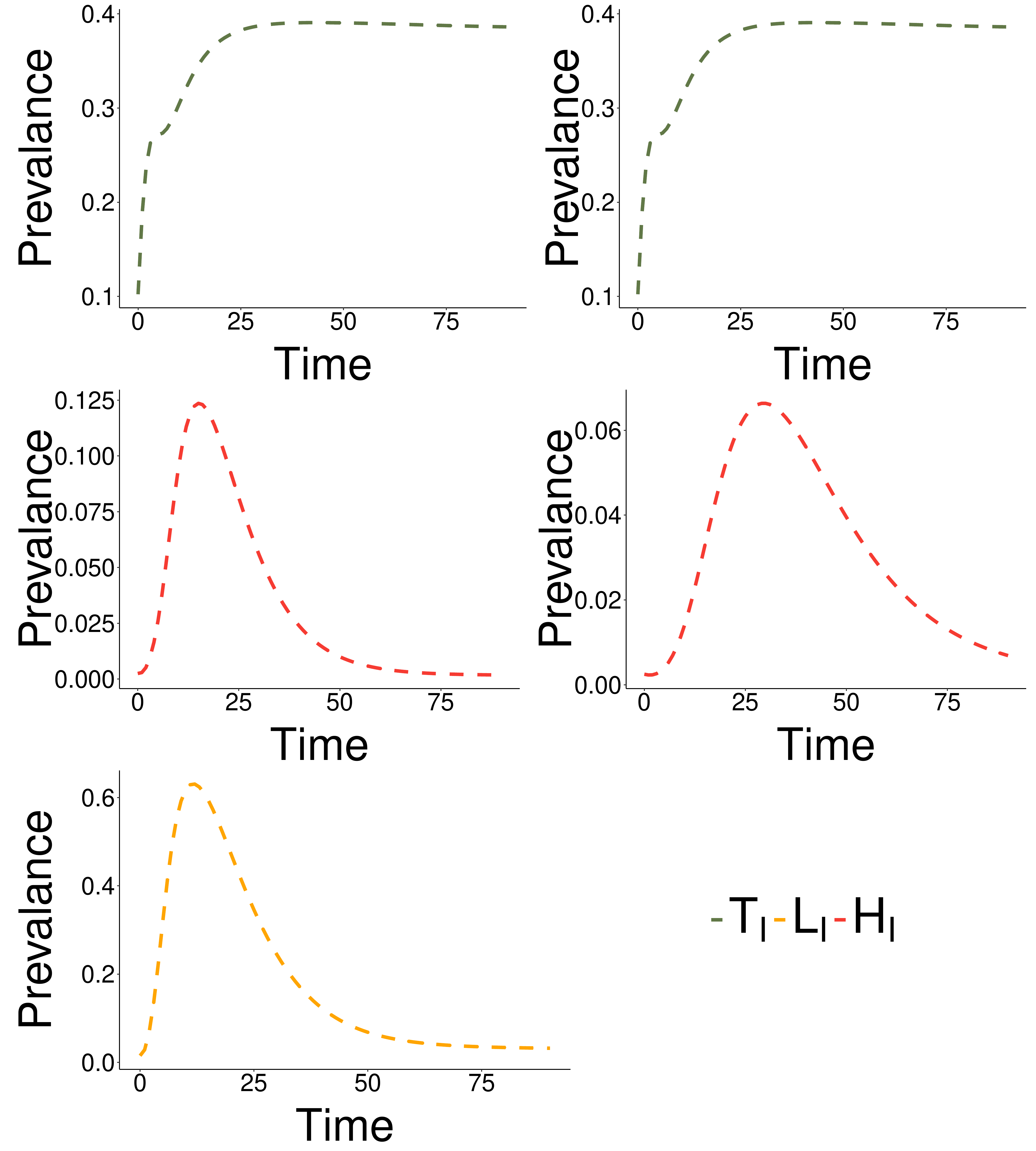}   
    \caption{
    Simulated infection dynamics of multi-vectors model what includes the nosocomial spread of CCHF. 
    CCHFV prevalence in the effective tick population, livestock and human populations while considering in this instance both infectious and removed host populations at the equilibrium point according to 
     \cite{10.1093/jme/tjy035}.
    CCHFV prevalence in the effective tick population and human populations without considering the livestock to human transmission path as described in section \ref{Tick-Human}
    }
 \label{Fig:Infection_Dynamics}
\end{figure}

We carry out a systematic examination to understand the long-term disease dynamics first in the model systems~\eqref{eq:1}, \eqref{eq:2}, \eqref{eq:3a} and in that of section \ref{Tick-Human} using the parameters given in Table \ref{table:2} and Table \ref{table:3}. 
Figure~\ref{Fig:Infection_Dynamics}  shows the infection dynamics incorporating the nosocomial spread and demonstrates the model system in section \ref{Tick-Human} without the transmission routes from livestock to human.

To demonstrate the infection profiles of  ticks ($T_I$), livestock ($L_I$) and humans ($H_I$) in our model \eqref{eq:1}, \eqref{eq:2}, \eqref{eq:3a} and that of section \ref{Tick-Human}, the initial conditions are mentioned in the Table  \ref{table:4}.
To represent our simulated data, we followed the modelling hypothesis as mentioned in \cite{10.1093/jme/tjy035}.
According to the the authors in \cite{10.1093/jme/tjy035}, simulated proportion of the  interacting populations that were infected at least once are calculated as the summation of infected and recovered or removed parts of the respective populations.

Few interesting things are observed in this simulation experiment.
The CCHFV prevalence in the  ticks do not show variations with time and it saturates, and the time average of the simulated prevalence in the ticks saturates around $40\%$.
Similarly, the simulated prevalence in the livestock and human populations show fluctuations. 
Simulated prevalence in livestock shows the highest prevalence around $62\%$ (bottom panel) and slowly it decreases.
We can claim that the prevalence in the livestock strongly depends on the ticks feeding on the host.
Simulated prevalence in humans in both cases (Figure \ref{Fig:Infection_Dynamics} middle panel) reveals the importance of the dissemination of CCHFV from livestock to humans.
According to the modelling experiment, when we include the CCHFV transmission from the livestock to humans, the simulated prevalence of CCHFV is around twice the same of the simulated prevalence 
of CCHFV without the inclusion of the disease transmission from livestock to humans.
This possibly shows the need for the proper care in handling of the infected livestock in certain geographical locations.
\begin{table}[H]
\centering
\begin{tabular}{||c c ||} 
 \hline
 \tiny{Variable} & \tiny{Initial Value}  \\ [0.5ex] 
 \hline\hline
  \tiny{$\mathit{T_S}$}& \tiny{30}  \\ 
  \tiny{$\mathit{T_E}$}& \tiny{10}  \\ 
  \tiny{$\mathit{T_I}$}& \tiny{10 } \\ 
  \tiny{$\mathit{L_S}$}& \tiny{970}  \\ 
  \tiny{$\mathit{L_E}$}& \tiny{20 } \\ 
  \tiny{$\mathit{L_I}$}& \tiny{10 } \\ 
  \tiny{$\mathit{L_R}$}& \tiny{0 } \\ 
  \tiny{$\mathit{H_S}$}& \tiny{99}  \\ 
  \tiny{$\mathit{H_E}$}& \tiny{0}  \\ 
  \tiny{$\mathit{H_I}$}& \tiny{1}  \\ 
  \tiny{$\mathit{H_R}$}& \tiny{0}  \\ 
 \hline
\end{tabular}
\caption{Initial values for the simulations.}
\label{table:4}
\end{table}

\section{Persistence-extinction boundary of CCHF}
\begin{figure}[H]
    \centering
    \includegraphics[width=0.4\textwidth]{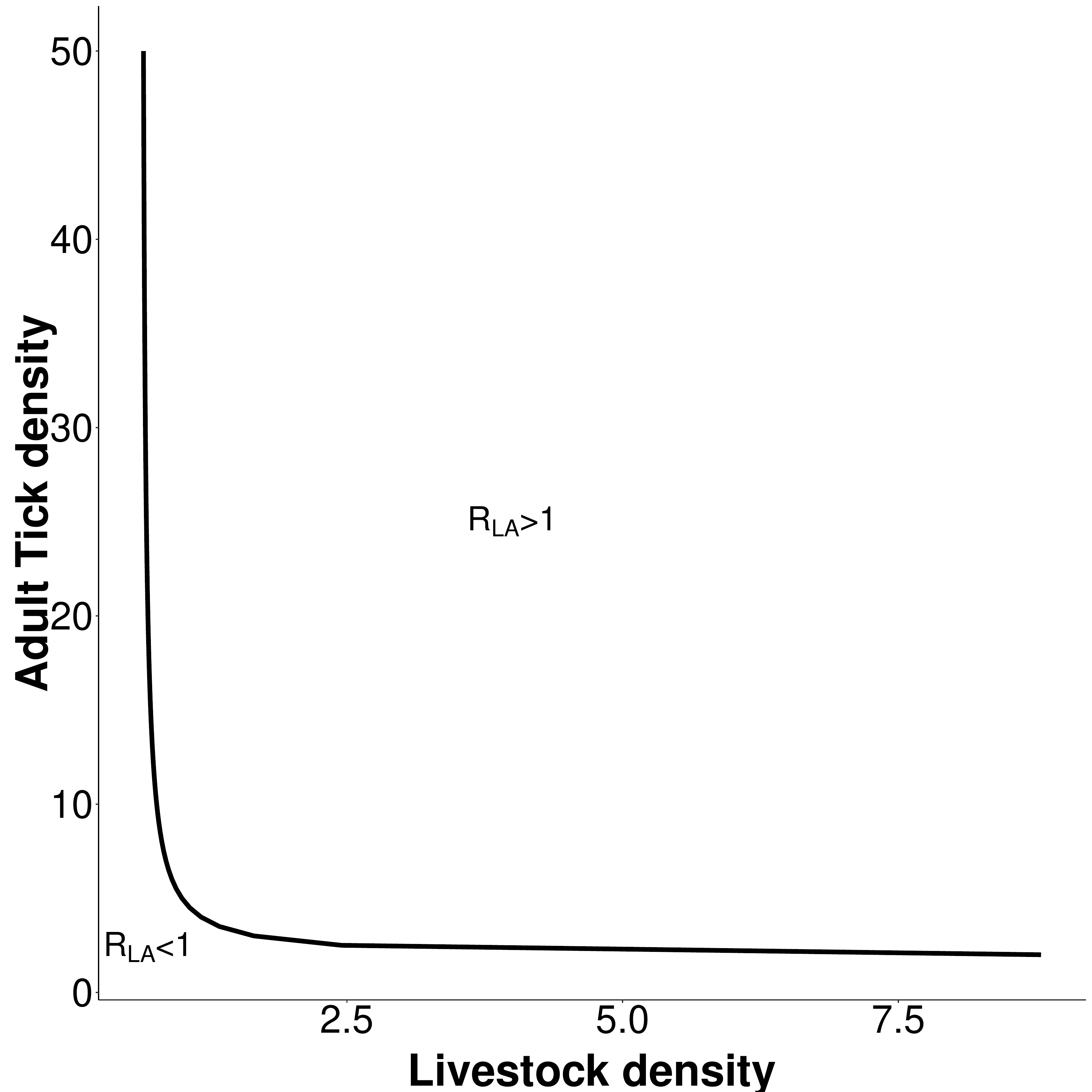}   
    \caption{Relationship between tick density and livestock density on the predicted area of CCHF persistence.}
 \label{Fig:TovsLo}
\end{figure}

After drawing the curve described by $R_{LA} = 1$ in \eqref{RTL}, we can observe from the Figure~\ref{Fig:TovsLo}, the required combinations of expected livestock densities that will lead
CCHF to persist and those that are not.
The curve for $R_{LA} = 1$ illustrates the possible expected cut-off point for CCHFV to persist.
Above this curve, CCHFV will persist, while it will die out below the curve.

\section{Control Strategies}
We use our model to analyse different control measures that can be employed by policy makers to decrease the number of human cases and the duration of outbreak situations.
With the presented multi-host model, the exploration of all possible control strategies is difficult to undertake.
This leaves us with the choice of aiming at particular host types only, such as vectors control, social distancing among humans, vaccinating livestock species etc. 
A new epidemiological metric named as \emph{Target reproduction number} ($\mathcal{T_S}$) is characterised to quantify the control measurements for  infectious diseases with multiple host types.
Therefore, in this situation the Target reproduction number ($\mathcal{T_S}$) is more useful compared to conventional $R_0$~\cite{HEESTERBEEK20073}.
This metric can be applied to study various control measures, when targeting subset of different types of hosts. 
Let us denote $\mathcal{K}$ \eqref{MatK} as following for convenience. 
\begin{equation} \label{KCont}  	
 \mathcal{K} = 
\begin{pmatrix}
K_{11} & K_{12} & 0 \\
K_{21} & 0 & 0 \\
K_{31} & K_{32} & K_{33} \\
\end{pmatrix}
\end{equation}
There are several options, through which we can effort to curb the bite of CCHFV-infected ticks.
Keeping the same notations as in \cite{Shuai2013}, the target reproduction number $\mathcal{T_S}$ with respect to the target set $\mathcal{S}$ is
defined as 
\begin{equation}\label{TS} 
\mathcal{T_S} = \bm{\rho}\left(\mathcal{K_S}\left(\mathcal{I}-\mathcal{K}+\mathcal{K_S}\right)^{-1}\right)
\end{equation}
where, $\mathcal{K_S}$ is the target matrix and defined as in \cite{Shuai2013} i.e. $[\mathcal{K_S}]_{ij}=K_{ij}$ if $(i,j)\in \mathcal{S}$ and $0$, otherwise.
$\mathcal{I}$ is the identity matrix  and $ \bm{\rho}$ is the spectral radius of the matrix.
Different disease control strategies are described below:\par
\emph{Livestock Sanitation:} The usage of acaricides is a common technique to lower the tick burden on the livestock. 
Then the target set is $\mathcal{S} =  \{(1,2),(2,1),(3,2)\}$, where the code types represent  the index pairs in \eqref{KCont}.
The type reproduction number targeting the host type 1 after employing \eqref{TS} is given by:
\begin{equation} 
\bm{\rho}
\begin{pmatrix}
0 & K_{12} & 0 \\
\frac{K_{21}}{1-K_{11}} & 0 & 0 \\
0 & K_{32} & 0
\end{pmatrix}
= \sqrt{\frac{K_{12}K_{21}}{1-K_{11}}}
\end{equation}
provided $K_{11}<1$.\par

\emph{Human Sanitation \& Isolation:} 
It is always advisable to wear proper clothing, while walking in the grazing field or to take precautionary measures when slaughtering livestock as well as during taking care of CCHFV-affected or -suspect patients.
Here the target set is $\mathcal{S} =  \{ (3,1),(3,2),(3,3) \}$. 
Target reproduction number $\mathcal{T_S}$ with respect to $\mathcal{S}$ (i.e., the type reproduction number targeting the host type $2$) is
\begin{equation} 
\bm{\rho}\tiny{
\begin{pmatrix}
0 & 0 & 0 \\
0 & 0 & 0 \\
-K_{31} {\left(\frac{1}{K_{11} - 1} - \frac{K_{12} K_{21}}{{\left(K_{11}
- 1\right)}^{2} {\left(\frac{K_{12} K_{21}}{K_{11} - 1} +
1\right)}}\right)} - \frac{K_{21} K_{32}}{{\left(K_{11} - 1\right)}
{\left(\frac{K_{12} K_{21}}{K_{11} - 1} + 1\right)}} & -\frac{K_{12}
K_{31}}{{\left(K_{11} - 1\right)} {\left(\frac{K_{12} K_{21}}{K_{11} -
1} + 1\right)}} + \frac{K_{32}}{\frac{K_{12} K_{21}}{K_{11} - 1} + 1}
& K_{33}
\end{pmatrix}
}\\
\end{equation}
$= K_{33}$.\par

\emph{Combined Control:} If we combine both the control options then our target set is $\mathcal{S} =  \{(1,2), (2,1), (3,1), (3,2), (3,3)\}$.
Target reproduction number $\mathcal{T_S}$ with respect to $\mathcal{S}$ is 
\begin{equation}
\bm{\rho}
\begin{pmatrix}
0 & K_{12} & 0 \\
\frac{K_{21}}{1-K_{11}} & 0 & 0 \\
\frac{K_{31}}{1-K_{11}} & K_{32} & K_{33}
\end{pmatrix}
= \max \{K_{33}, \sqrt{\frac{K_{12}K_{21}}{1-K_{11}}}\}
\end{equation}
\emph{Isolation:}
It is difficult to prevent or control the CCHFV infection cycle in livestock and ticks, as the tick–animal–tick cycle usually goes unnoticed, and CCHFV infection in livestock is not evident due to the lack of clinical signs of infection.
Moreover, the abundance of tick vectors is widespread and large in numbers, which requires an efficient tick control strategy.
This may be possible mainly in structured livestock farms.
In those farms, where tick control may not be possible due to economic constraints~\cite{ Atif2017, BANNAZADEHBAGHI2016634}, only isolation could be a realistic option.
In this situation the target set is $\mathcal{S} =  \{(3,3)\}$.
\begin{equation}
\bm{\rho}
\begin{pmatrix}
0 & 0 & 0 \\
0 & 0 & 0 \\
K_{33} {\left(\frac{K_{21} {\left(\frac{K_{12} K_{31}}{K_{11} - 1} -
K_{32}\right)}}{{\left(K_{11} - 1\right)} {\left(\frac{K_{12}
K_{21}}{k_{11} - 1} + 1\right)}} - \frac{k_{31}}{k_{11} - 1}\right)}
& -\frac{{\left(\frac{K_{12} K_{31}}{K_{11} - 1} - K_{32}\right)}
K_{33}}{\frac{K_{12} K_{21}}{K_{11} - 1} + 1} & K_{33}
\end{pmatrix}
= K_{33}
\end{equation}

It is interesting to notice from the mathematical perspective that the efforts required to eradicate the disease are  same for both \emph{Human Sanitation \& Isolation} and only for \emph{Isolation}.
This can be attributed to the fact that the latter case is a subset of the former control method.
\par
A compelling choice to reduce the potential risk of CCHFV in the livestock is to keep also chickens as they pick the ticks that is responsible for the transmission of CCHFV~\cite{KASI2020101324}.

\section{Sensitivity Analysis}
In this section we carry out a sensitivity analysis of the model parameters to the model output to deduce the important parameters that may help 
to control the CCHF infection.
It can be defined as the treatise of how uncertainty in the output of a mathematical model can correspond to different sources of uncertainty in the model input parameters~\cite{Iooss2017}.
It is a technique that systemically varies the model input parameters and thus helps to determine their effects on the model output.
\subsection{Model Sensitivity Analysis}
Given the large number of model parameters, it is instructive to obtain those model parameters that have the greatest influence on CCHF transmission.
We therefore perform the sensitivity analysis through computing the Partial Rank Correlation Coefficients (PRCC) with 1000 simulations per run for each of the model input parameter values sampled by the Latin Hypercube Sampling (LHS) scheme.
This method has an assumption that there is a monotonic relationship between the model input parameters and the model outputs.
Here, we consider the cumulative human cases of CCHF occurring during a simulation experiment as the model output of interest without the human-to-human spread.
This approach has the advantage that it captures the effects of model parameters on both, the persistence of CCHF and the overall impact of CCHF outbreaks over time. 
The sign of the PRCC values depicts the qualitative relationship between the model input parameters and the model output of interests.
A positive PRCC values means that while the corresponding model input parameters increases, the model output will also increase and on the other hand, a negative PRCC value suggests
a negative correlation between the model in- and output \cite{6088374}.
Values near zero indicate little effect on the model output.
\begin{figure}[H]
    \centering
    \includegraphics[width=0.7\textwidth]{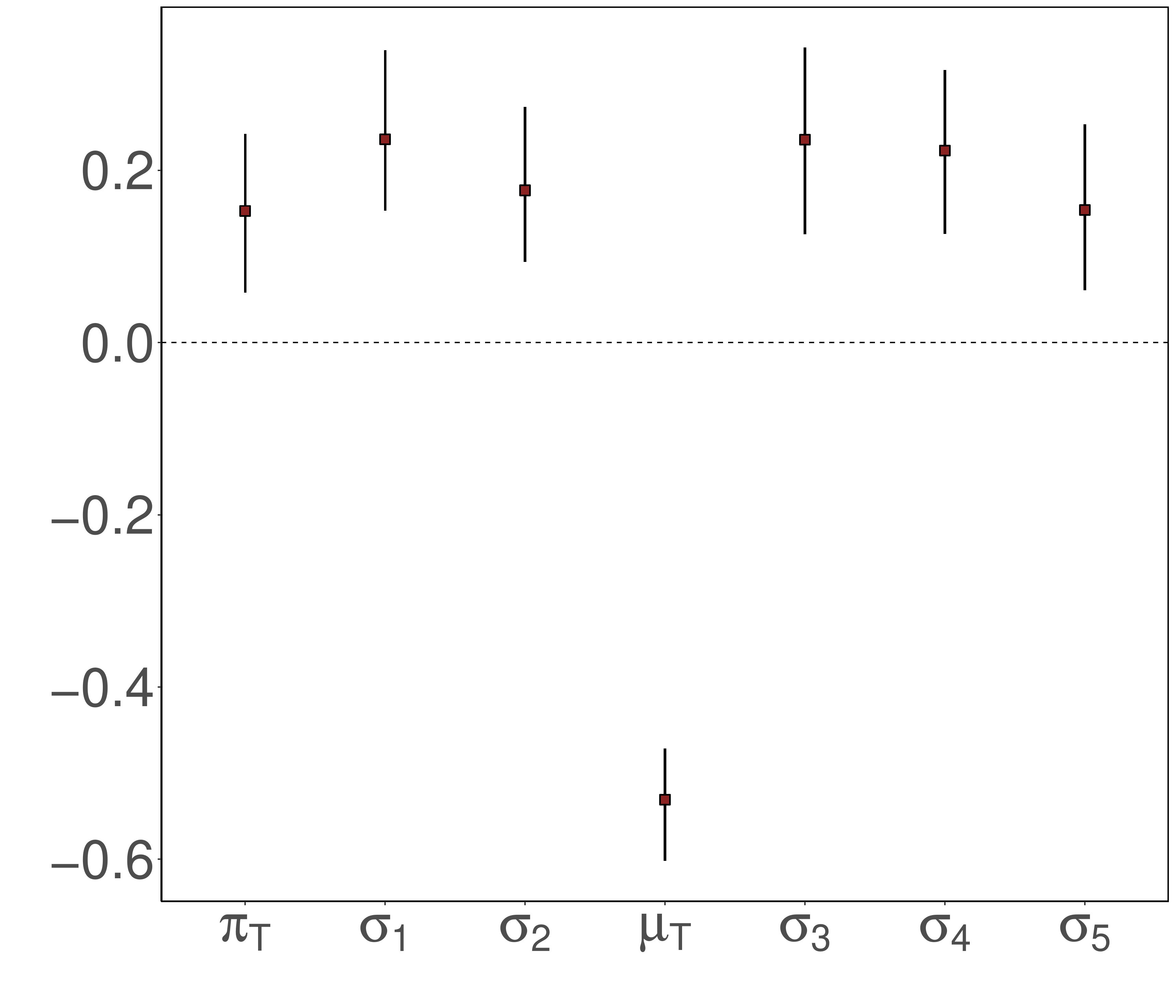}     
    \caption{PRCC Analysis}
    \label{Fig:Sens1}
\end{figure}
We can observe in Figure~\ref{Fig:Sens1} that the mortality of the infected  ticks ($\mu_T$) shows a strong negative correlations with the cumulative incidences of CCHF in humans, whereas 
the effective transmission between tick and livestock ($\sigma_3$, $\sigma_1$),  effective transmission amongst the ticks through co-feeding ($\sigma_2$), the ticks birth rate ($\pi_T$),
the effective transmission between ticks and humans ($\sigma_4$) and effective contact rate between livestock and humans ($\sigma_5$) show the strong positive correlations with the model output.\par
Therefore, after using the results of sensitivity analysis we can conclude that the parameters with the strongest influence on the cumulative incidences of CCHF in humans are $\mu_T$, $\pi_T$
and $\sigma_i$, where ($i= 1, 2, 3, 4, 5$).
Cumulative incidences of CCHF in human increase with the increase in $\pi_T$ and $\sigma_i$, where ($i= 1, 2, 3, 4, 5$) and decrease with $\mu_T$.
To perform the sensitivity analysis we use the \textit{sensitivity}~\cite{sensitivity} package and for the LHS scheme we utilise the \textit{lhs} \cite{lhs} package in R~\cite{R}.

\subsection{NGM Sensitivity Analysis}
While sensitivity analysis of the model parameters gives us insights into the dynamics of the model, we can also measure the  impact of parameters on $R_0$ directly.
Therefore, we compute the sensitivity and elasticity of NGM directly.
Sensitivities and elasticities are measures of how infinitesimal changes in individual entries of a stage-structured population matrix will affect the population and the quantification of projection results on the parameters. 
After noticing $R_0$ \eqref{RH,RLA} as a function of $\mathcal{K} [X, Y]$ \eqref{KInter}, we denote 
\begin{equation}\label{Sens1}
\mathcal{S}_{X,Y} = \frac{\partial R_0}{\partial \mathcal{K} [X, Y]}
\end{equation}
as the sensitivity of $R_0$ and 
\begin{equation}\label{Elasti1}
\mathcal{T}_{X,Y}=\mathcal{K} [X, Y]\frac{ \partial [\ln R_0]}{\partial \mathcal{K} [X, Y]}
\end{equation}
as the elasticity of $R_0$.

Following \cite{POLO2018119}, we perform sensitivity and elasticity analyses of \eqref{KInter} in R~\cite{R} using the package \textit{popbio} \cite{popbio} which is an R version of the Matlab code for the analysis of matrix population models illustrated in \cite{10007469608}.
\begin{figure}[H]
    \centering
    \includegraphics[width=1\textwidth]{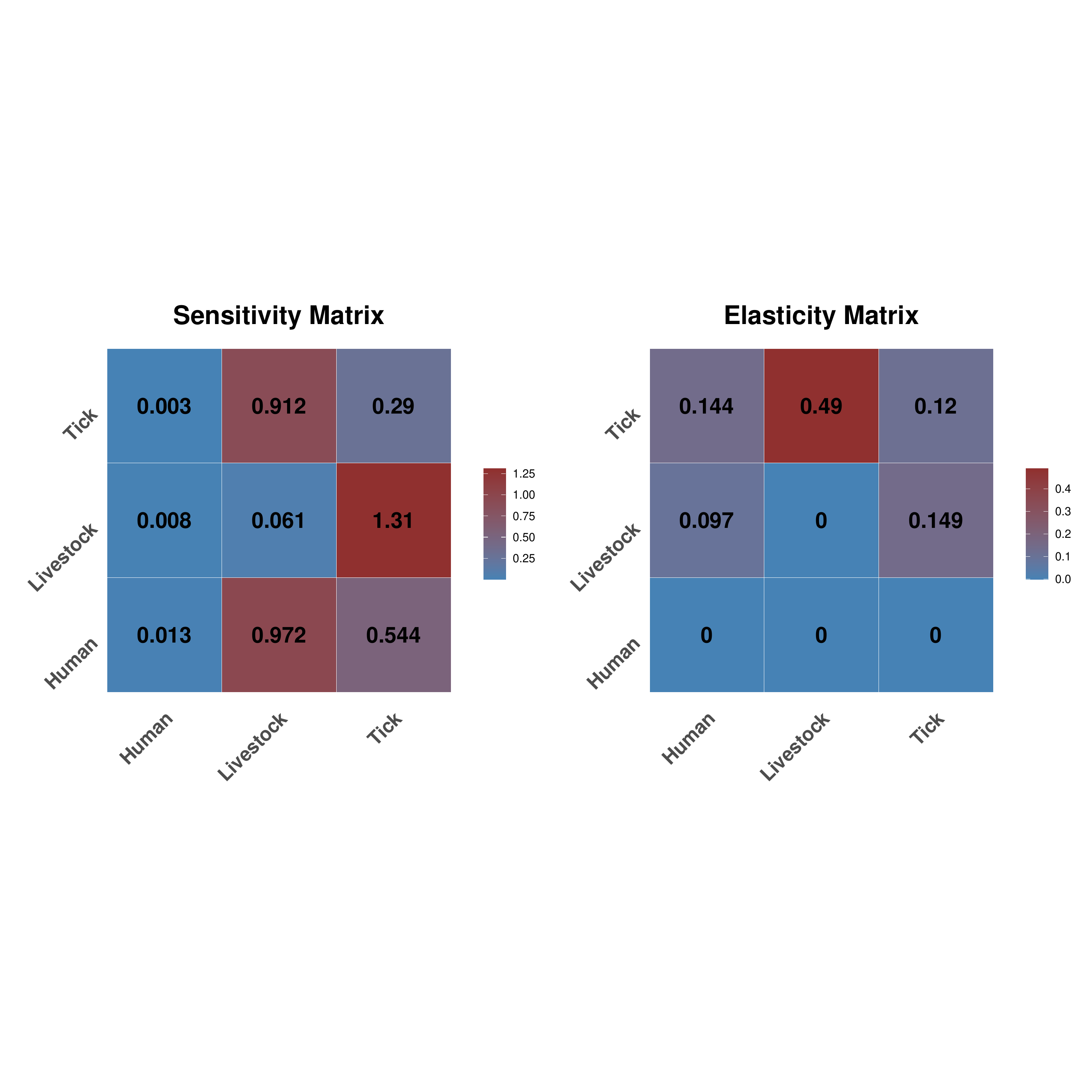}     
    \caption{(a) Sensitivity and (b) elasticity matrices for ~\eqref{KInter}}.
    \label{Fig:NGMSensi}
\end{figure}
From Figure~\ref{Fig:NGMSensi}, it is noticeable that the value of $R_0$~\eqref{RH,RLA} is sensitive to the changes in $K_{12}$, $K_{21}$, $K_{32}$ and $K_{31}$ of the elements from the matrix $\mathcal{K}$~\eqref{KInter}.
These correspond to the numbers of CCHFV-infected  ticks attached to an infected livestock animal and the numbers of infected livestock animals infested by an infected  tick, followed by the number of infected humans bitten by an infected  tick and the number of infected humans produced by a single infected livestock animal.
It is also interesting to note that the number of infected  ticks produced by a single infected tick through co-feeding ($K_{11}$) is also a sensitive parameter in the model.\par
Elasticities are actually \textit{proportional sensitivities} (\cite{LESNOFF2003945}) which measure the proportional change in $R_0$~\eqref{RH,RLA}, given an infinitesimal one-at-a-time proportional change in the  elements of the matrix $\mathcal{K}$~\eqref{KInter} with the assumption that $\mathcal{K}$ is growing or decreasing at a constant rate~(\cite{10007469608}).
Figure~\ref{Fig:NGMSensi} shows the elasticity of $R_0$ with respect to the matrix elements $\mathcal{K} [X, Y]$.
The elasticities of the matrix elements $K_{11}$, $K_{12}$, $K_{13}$ and $K_{23}$ adds up--to approximately $90.3\%$.
Simple interpretations of elasticities can provide a metric that illustrates the relative importance of the disease cycle, both within and between the host-tick populations.
However, we emphasise again that elasticities are formulated on infinitesimal, one-at-a-time changes, with the information that the multiple changes are additive and that the effects of the changes of  $\mathcal{K} [X, Y]$ are assumed to be linear~\cite{10007469608}.
The interaction between the infected tick and livestock are the prime factor driving the CCHF cycle.

\section{Model Fitting}
Mathematical models of disease dynamics and control have ample applications: to understand the hidden functioning of a mechanism, to simulate experiments prior to perform them etc. 
Some of the associated parameters can be found by conducting experiments or in the literature. However, for our proposed model, we lack the values and the distributions of different parameters. 
To validate the robustness of our ODE model \eqref{eq:T1a}, \eqref{eq:H3b}, we have fitted it to the actual CCHFV incidence data from six different countries. 
To perform this data fitting process, we have used the MATLAB\textsuperscript{\textregistered} \cite{MATLAB:2019} differential equation solver \textit{ode45} to approximate the solution for a trial set of parameter values with the fixed initial condition. 
We use Matlab functions \textit{fminsearch} and \textit{lsqcurvefit}.

\begin{figure}[h]
\centering
\subfloat[Subfigure 1 list of figures text][Afghanistan]{
\includegraphics[width=0.28\textwidth]{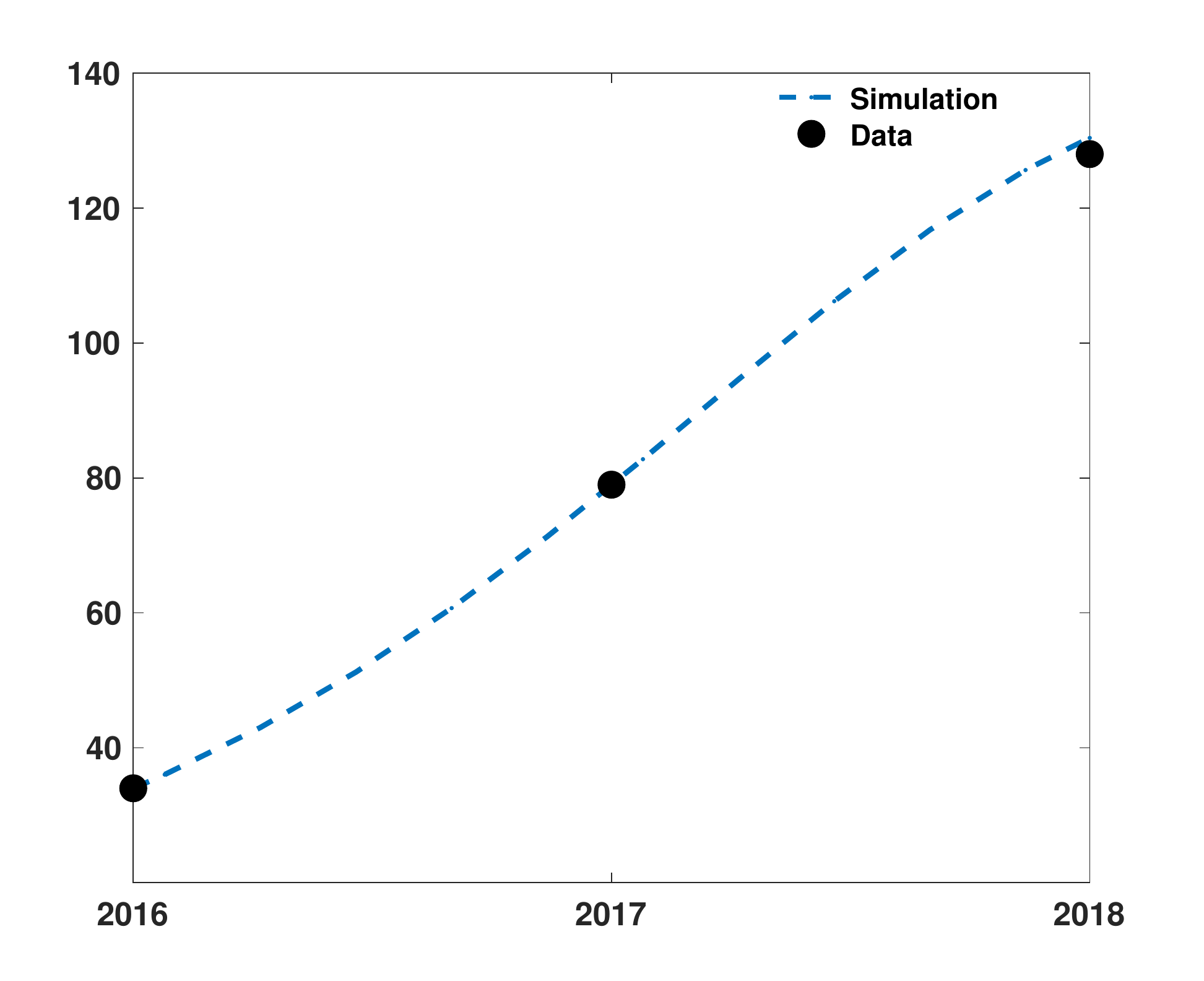}
\label{fig:subfig1}}
\subfloat[Subfigure 2 list of figures text][Bulgaria]{
\includegraphics[width=0.3\textwidth]{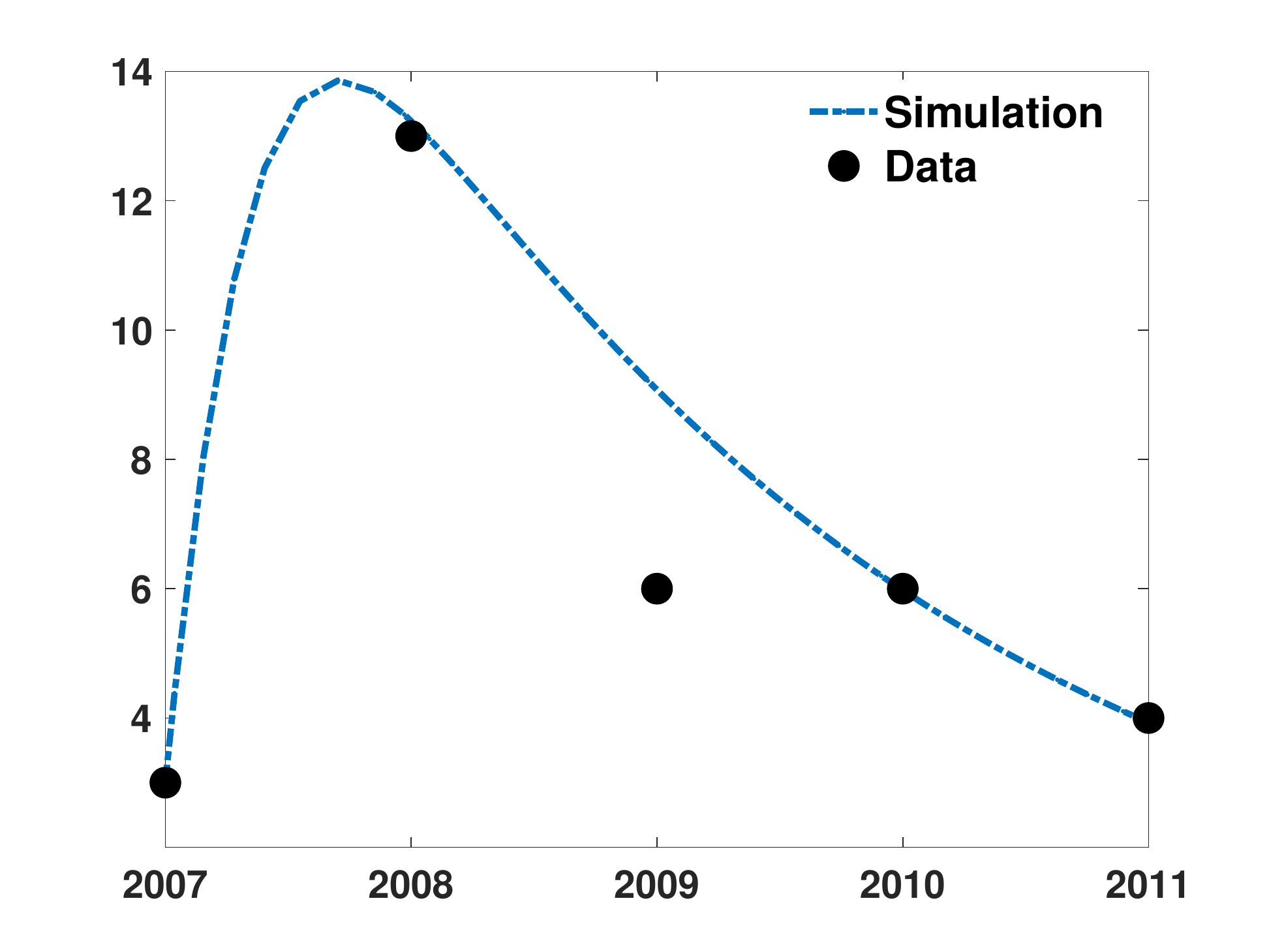}
\label{fig:subfig2}}
\subfloat[Subfigure 3 list of figures text][Kosovo]{
\includegraphics[width=0.3\textwidth]{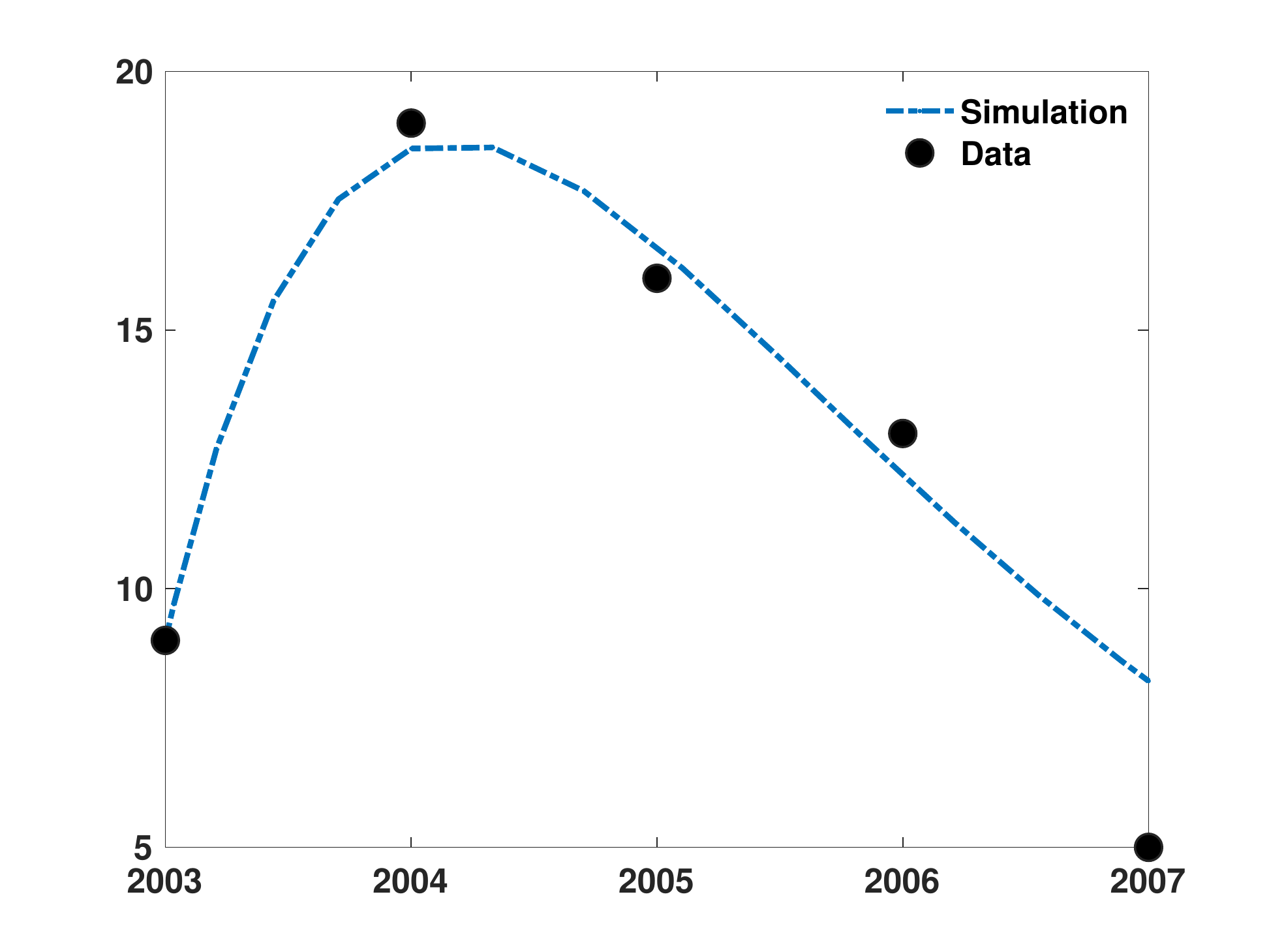}
\label{fig:subfig3}}
\qquad
\subfloat[Subfigure 4 list of figures text][Turkey]{
\includegraphics[width=0.3\textwidth]{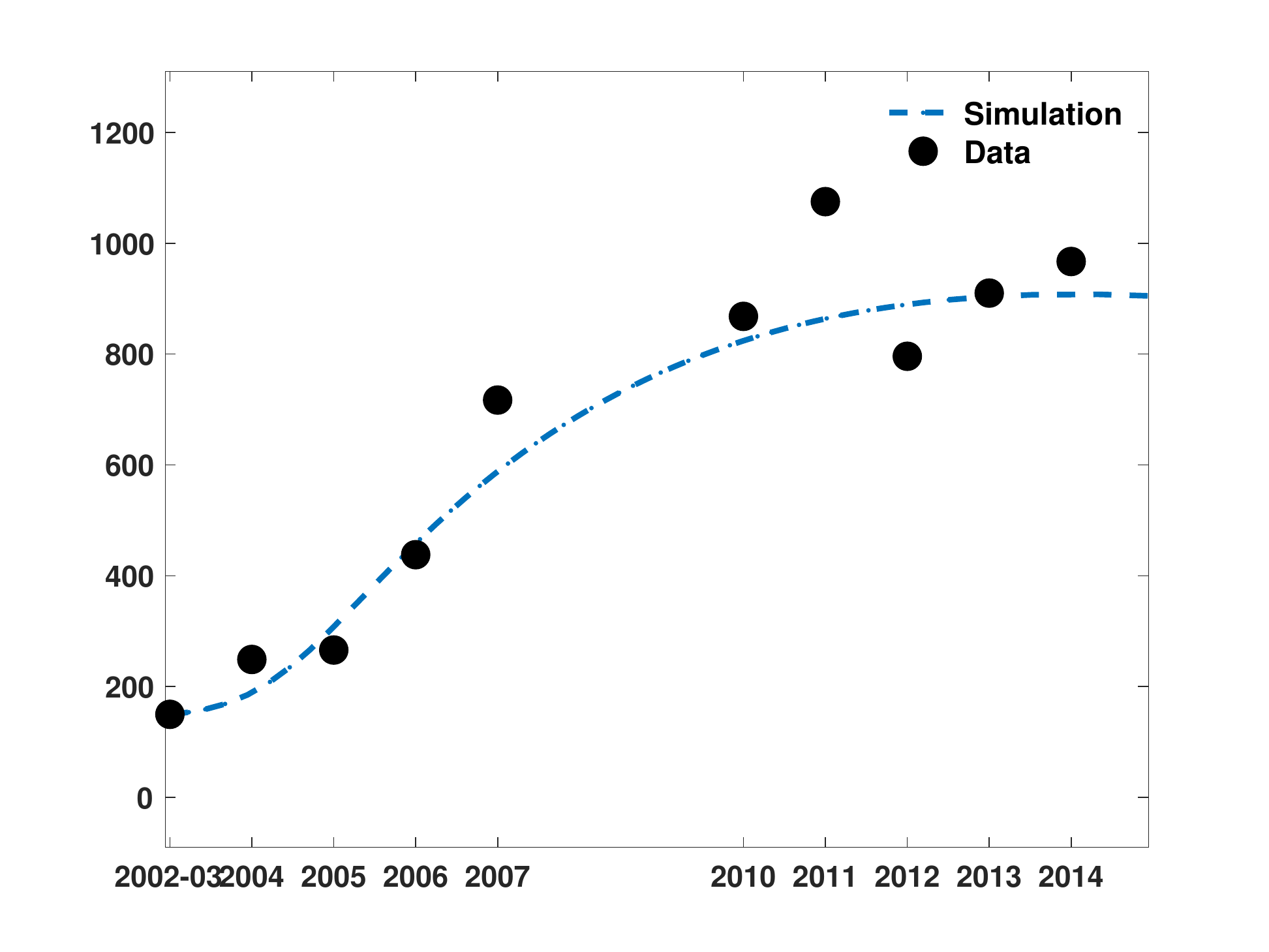}
\label{fig:subfig4}}
\subfloat[Subfigure 5 list of figures text][Pakistan]{
\includegraphics[width=0.3\textwidth]{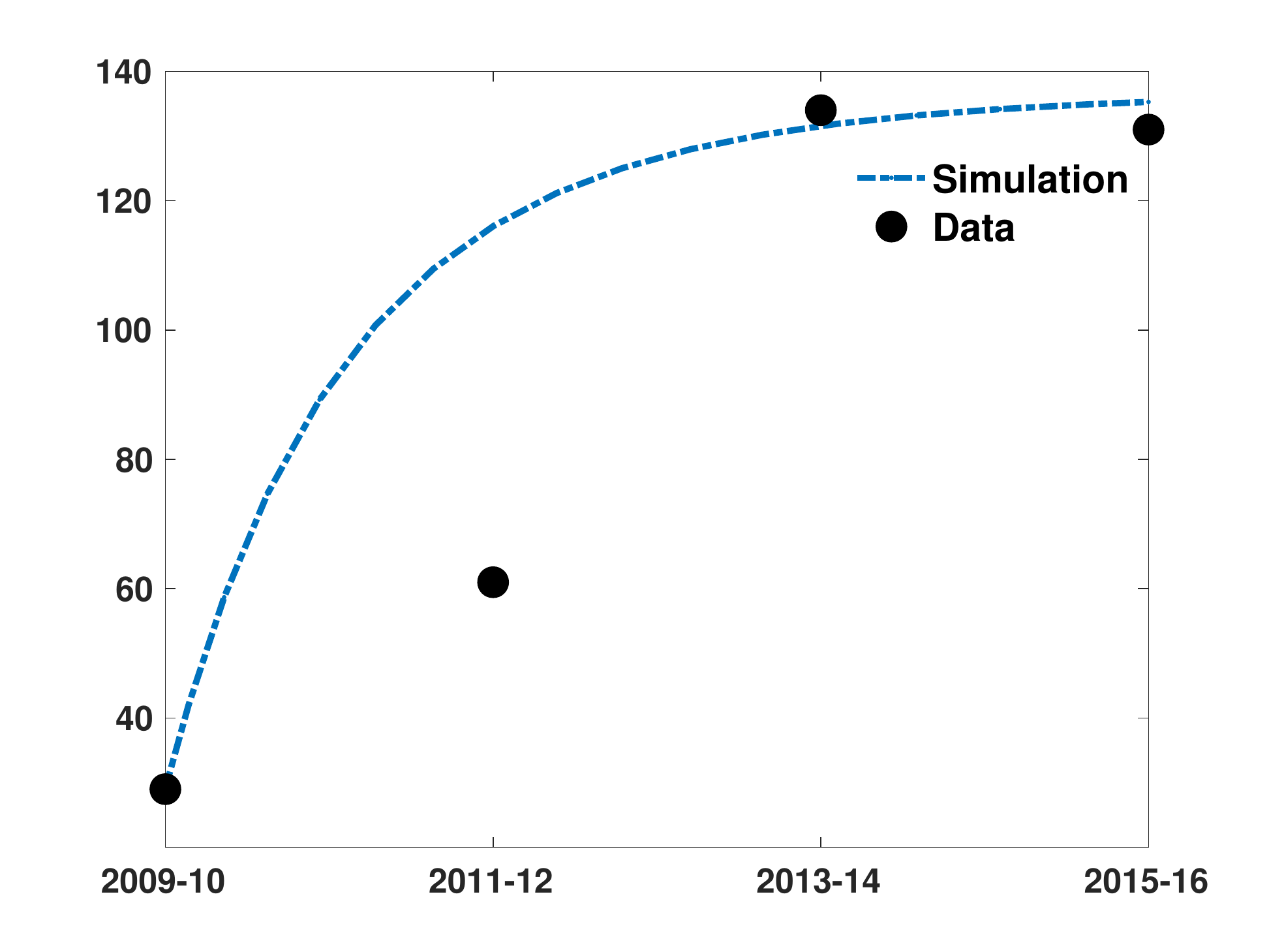}
\label{fig:subfig5}}
\subfloat[Subfigure 6 list of figures text][Iran]{
\includegraphics[width=0.3\textwidth]{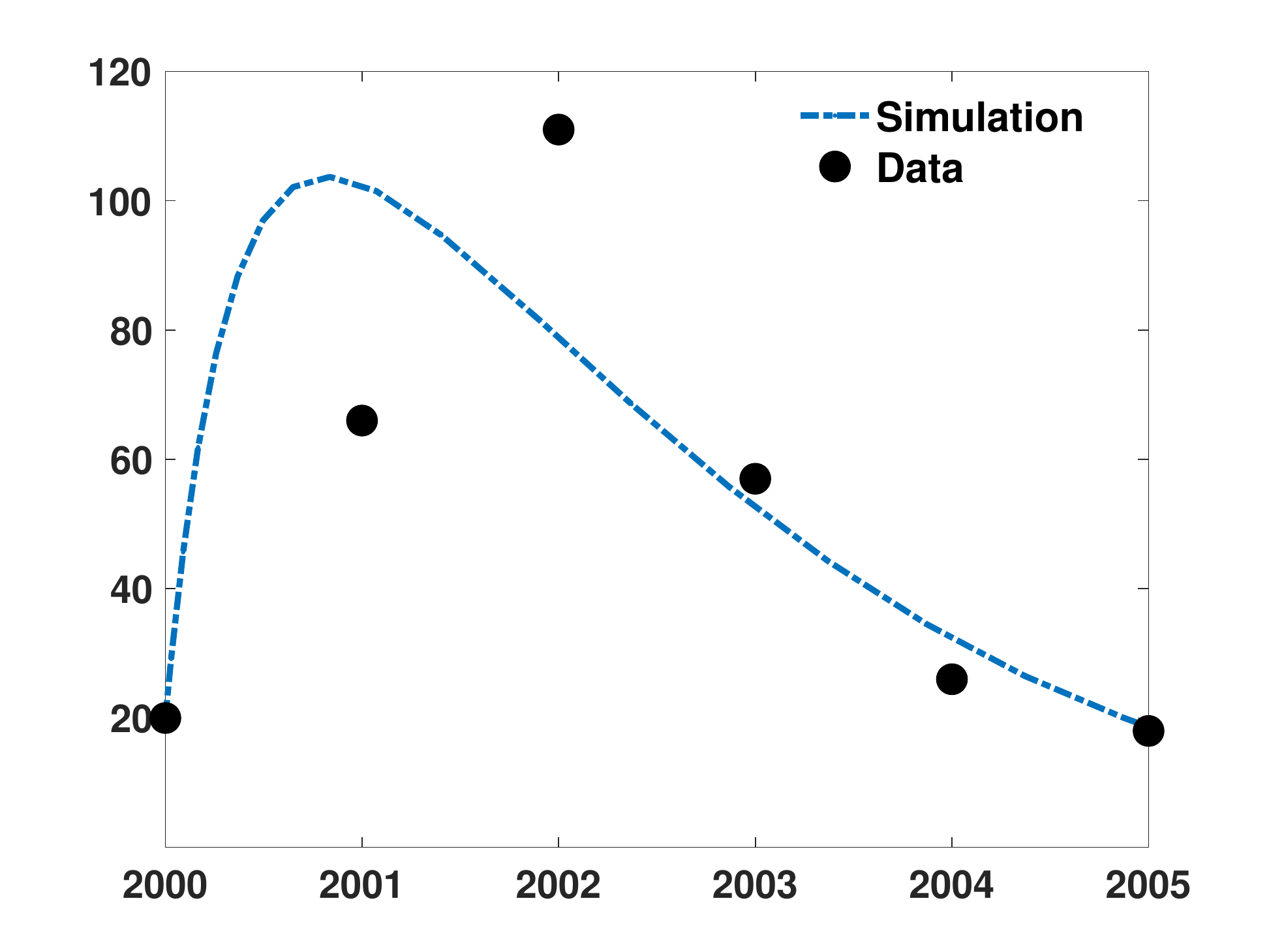}
\label{fig:subfig6}}

\caption{The comparison between the reported human CCHFV cases in Afghanistan, Bulgaria, Kosovo, Turkey, Pakistan and Iran and the simulation of $H_I(t)$ from the model. 
Data from \cite{ECDC,10.1371/journal.pone.0110982, Atif2017, WHO, Hege, PBS, FAO1, SOK, USDA, FAO, Kam, FAO2}}.
\label{fig:globfig}
\end{figure}

Fitting~\eqref{eq:T1a}, \eqref{eq:H3b} to real incidence data is important for modelling, as well as to have certain basic parameters, around which we can vary and run simulations to explore various disease spread scenarios.
The numerical simulation of human CCHFV cases in different countries is shown in Figure~\ref{fig:globfig}.
The increased awareness towards the perils of CCHFV may have helped to decrease the cases for cases in Bulgaria, Iran and  Kosovo, but in other countries, it appears that this is not the case. 
Moreover, our fitted model simulations (Figure~\ref{fig:subfig1}, \ref{fig:subfig4} and \ref{fig:subfig5}) demonstrate that,  given the current trend of the CCHFV cases in Afghanistan, Pakistan and Turkey,  the number of human CCHFV cases will keep on increasing  in future. 
Therefore, if no further effective prevention and control measures are taken, the disease will not vanish.
Visualisations of the fitted parameters are included in the Supplementary Information (SI).
\par

\section{Multidimensional Scaling Analysis }
\begin{figure}[H]
\centering
\subfloat[Subfigure 1 list of figures text][]{
\includegraphics[height=60mm,width=60mm]{output11-eps-converted-to.pdf}
\label{Fig:CosineSim}}
\subfloat[Subfigure 4 list of figures text][]{
\includegraphics[height=60mm,width=60mm]{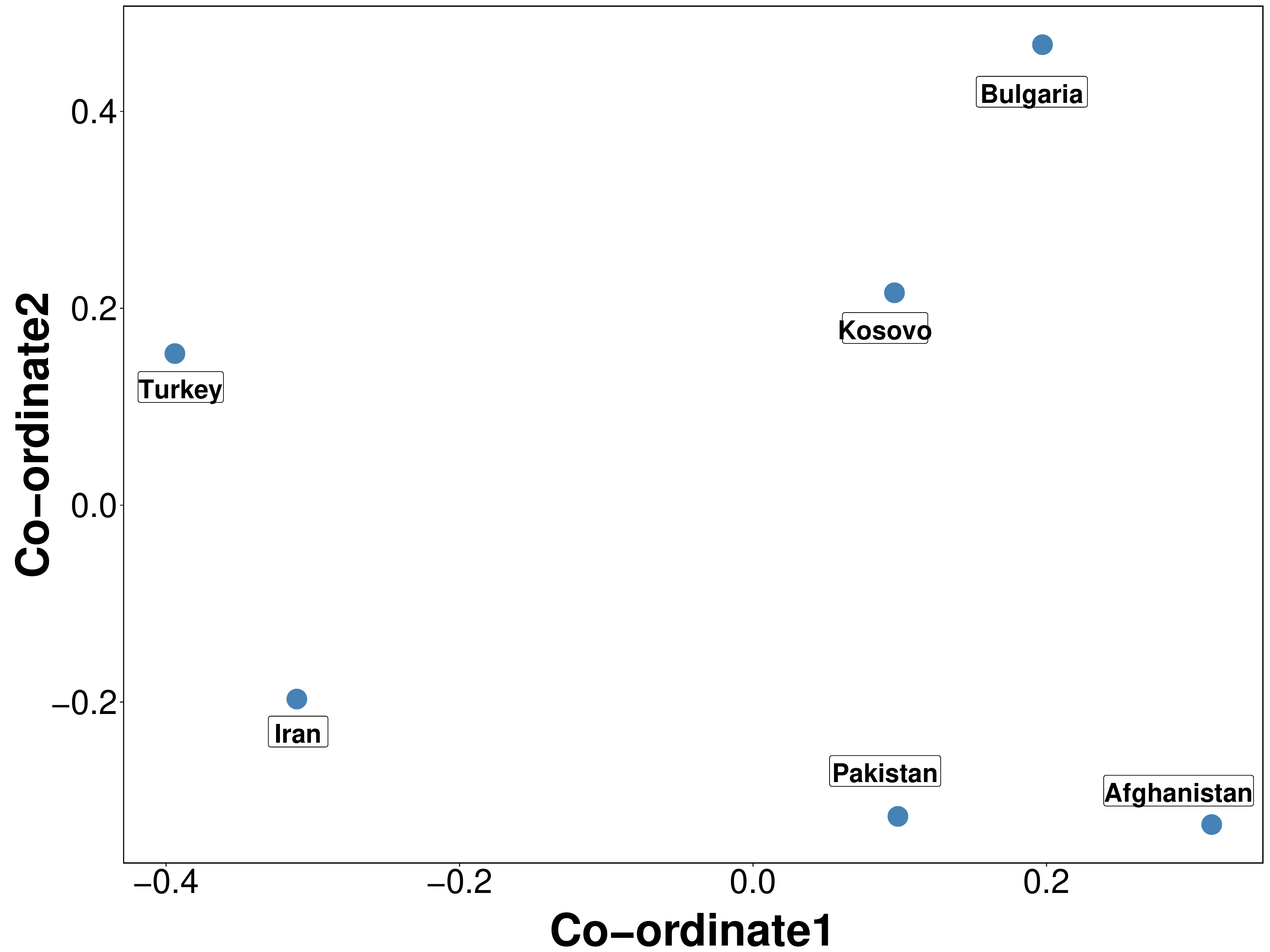}
\label{Fig:MDS}}
\caption{ Differences among the parameter sets of the considered countries.
(a) Cosine similarity of the distance matrix (b) Spatial embedding of distance matrix.}
\label{Fig:ApprParms}
\end{figure}

A central question is: How much do the found parameters differ for the countries of concern?
In order to find an answer, we endeavour to find the cosine similarity index amongst the fitted transmission coefficients to inquire about the circulation of CCHFV transmission.
Cosine similarity index is a gauge of similarity between two non-zero vectors that measures the cosine of the angle between them.
Afterwards, we employ multidimensional scaling to visualise the level of similarity or dissimilarity of the fitted transmission parameters of concerned countries. 
We then interpret the findings about the pairwise cosine similarity index among the set of countries of consideration mapped into an abstract Cartesian space.
It yields a useful tool to quantify, how similar infection profiles of different countries will be in terms of transmission parameters. From Figure~\ref{Fig:CosineSim} it is evident that the disease transmission parameters are not equal. We can clearly observe that Afghanistan is the most affected country and has the highest disease transmission coefficient (Figure~\ref{Fig:CosineSim}). Afghanistan, Pakistan, Iran and Turkey are closer to each other as compared to the Balkan countries Bulgaria and Kosovo. 
We are also interested to learn about the infection profiles of different countries of concern while quantifying the burden of CCHFV dissemination by distinct transmission routes.
With the above said purpose, we employ multidimensional scaling to infer about the high-dimensional parameter space of the fitted transmission parameters. 
Figure~\ref{Fig:MDS} depicts a $2$ dimensional spatial embedding reflects the cosine distances of the parameters. 
It is clear that infection profiles of South Asian countries differs significantly from the Balkan and Middle Eastern countries.
Perhaps this information might be helpful to policy makers to adapt CCHFV control measures.

\section{Discussion \& Conclusion}
In recent years, a surge in CCHF cases has been observed, with an expansion beyond its historical spatial range~\cite{vetsci5020060, BARZON201838}.
Several factors possibly influence the geographic expansion of CCHFV, such as environmental change and movement of hosts carrying \textit{Hyalomma} ticks to new geographical areas.
In past, studies on tick-borne diseases e.g. Lyme disease~\cite{Lyme}, Tick-borne encephalitis (TBE)~\cite{Gaff2007} etc. have been conducted, however, limited studies have been reported on mathematical emphasise on the CCHFV~\cite{doi:10.1080/17513758.2015.1102976, Cooper2007, Hoch2016,10.1093/jme/tjy035}, as to our knowledge.
In our study, we construct CCHF transmission dynamics models (deterministic ODE models), which include the interactions amongst the \textit{Hyalomma} ticks, livestocks and human. 
We further extend our basic multi-vectors model by including the nosocomial spread of CCHF to analyse the effects of human-human transmission on the disease dynamics.\par
We analyse a new mathematical model which includes multiple transmission avenues of CCHFV and compare it with the model that only incorporates the transmission of the CCHFV
virus through direct contact with livestock and ticks.
Our model shows the importance of including of CCHFV transmission through co-feeding and its sustainability in  tick population.
Co-feeding escalates the value of the basic reproduction number $R_0$ and \eqref{R_LA} quantitatively depicts the increased value.   
With the assumed parameter list, our model predicts that the reduction of $18\%$ of the contact rate between the ticks and the livestock can reduce the value of the basic reproduction number.
When we consider only the tick-human model in Section \ref{Tick-Human}, we observe a reduction of $37\%$ in the contact rates which can help in reducing the value of the basic reproduction below $1$.
Simulations show that livestock has a significant role in disease transmission compare to tick-human model in the multi-vectors model. 
There is an increase of approximately $65\%$  in human CCHF cases due to contact with infected animal blood etc.
We propose that these additional pathways do not only increase the basic reproductive number of CCHF, along with influence in the infection profiles of the multi-vectors system, but have also a dominant role in CCHF control measures.
Computation of $R_{0}$ gives us the necessary tool to investigate different strategies to control the spread of vectors-borne diseases. 
Results from our simulations suggest the importance of the birth rate into the susceptible adult tick population in the disease dynamics and hence the influence of the rodent density implicitly. 
Linear growth in the tick population is reflected in the basic reproduction number (both in $R_{LA}$ and $R_{T}$) as it shows a linear relationship
Freshly recruited livestock provides the necessary number of susceptible livestock which increases the value of the basic reproduction number.
We also explore the influence of the effective transmission amongst ticks on the basic reproduction number.
We observe a linear trend among $\sigma_2$, $R_{LA}$ and $R_{T}$.
This may show that co-feeding of ticks alone may enable CCHFV transmission and increase the risk of transmission to humans substantially. 
The transmission rate of CCHFV from the  tick to livestock is an important parameter and the increase in effective contact between questing adult ticks and the livestock species increases the basic reproduction number in a nonlinear way.
\par
Our mathematical model is novel in the sense that it incorporates multiple transmission routes.
Some of the results are in accordance with the findings of previous studies. 
For example, the analytical expression of $R_{0}$ in \eqref{R0} is similar to the one in~\cite{doi:10.1080/17513758.2015.1102976} including the stability conditions.
Moreover the basic reproductive number for CCHF increased by the factor mention in ~\eqref{R_LA} through co-feeding, the threshold quantities are comparable in~\cite{doi:10.1080/17513758.2015.1102976}.
The numerical approximations of $R_{0}$ and in~\cite{doi:10.1111/j.1461-0248.2009.01378.x} are comparable.
The magnitude of the fitted parameters are in accord with previous findings \cite{Cooper2007,doi:10.1111/j.1461-0248.2009.01378.x}.
Simulated CCHFV prevalence in the tick and the livestock populations are of similar magnitude through our static model is of \cite{10.1093/jme/tjy035} without the influence of meteorological factors.
\par
Some authors \cite{Hoch2016, doi:10.1111/j.1461-0248.2009.01378.x, 10.1093/jme/tjy035} perform also a sensitivity analysis by analysing the effects of parameter values on $R_{0}$, but our study is different, as we analyse the sensitivity on another important epidemiological quantity, i.e. the total number of infected humans. 
Furthermore, sensitivity analysis of this quantity has not been examined elsewhere. 
Similar to \cite{Hoch2016, 10.1093/jme/tjy035}, our findings show that host density, duration of infection and the immune responses are sensitive parameters.
\par
While CCHFV has been studied in~\cite{doi:10.1080/17513758.2015.1102976} with livestock as the primary host, the systematic exploration of the model parameters and the mathematical illustrations of different control measures have not yet been fully explored.
For the tick-borne diseases, the use of acaricides is the primary treatment and it mainly target the tick population, but in some poor regions, this may not be feasible. 
However, we should bear in mind that the profuse usage of acaricides may have a detrimental effect on the environment.
In~\cite{doi:10.1080/17513758.2015.1102976}, the authors explore this option. 
However, alternatives to using acaricides have been proposed, e.g. keeping chickens together with other livestock, because chickens eat ticks and may therefore reduce the risk of exposure to CCHFV~\cite{KASI2020101324}.
\par
We also methodically explore all possible ways to curb CCHF cases in humans.
We find that the mechanism of CCHFV dissemination varies in different endemic countries.
Our results show that human sanitation and isolation are effective ways to reduce the CCHF cases in humans along with the acaricide treatment as mentioned in~\cite{Leb}.
Spatially embedded multidimensional scaling along with the cosine similarity index amongst the transmission parameters give us the clue that, the burden of CCHFV transmission differs from country to country. 
However, the control mechanisms need to be adapted to the specific situation in a country. 
The potential effects of measures can first be simulated using our model and any measures adapted accordingly, if the outcome in the model warrants this.
\par
Just like other modelling endeavours \cite{doi:10.1080/17513758.2015.1102976, Cooper2007} on CCHF, our model has several limitations.
Our model, like many others is based on assumption, where knowledge and parameters were missing. 
Variables were parametrised with values from the literature, which may be accurate or not, generally applicable or not. 
Therefore, the simulations conducted with our model are only meant for demonstration purposes. 
We recommend to parametrise the model for the specific situation, if it is used to plan or evaluate control measures. 
\par
Future studies with this model should include the proper investigation of the data related to CCHF and systematic explorations of the parameter space to find the necessary paths to reduce the disease prevalence effectively.
For simplicity, we have not included transovarial CCHFV tranmission~\cite{10.1309/LMN1P2FRZ7BKZSCO}.
Other investigators~\cite{Abbas2017} observe seasonality in human incidence and a dependence on ambient temperatures. 
These factors are not included in our approach. 
Ideally, our model may encompass such a barrier, if the effect of the seasonality and the dependence of environmental factors are included when transmission dynamics are modelled. 
The movement of animals and the migration of humans are also not included in this study, although these may be important variables.
According to~\cite{Atif2017}, even in urban areas of Pakistan, the risk of transmission is higher during the time of Eid-ul-Azha, when Muslims slaughter livestock animals. 
Periodic transmission risks may be included in our model, however, if appropriate.
\par
Despite the limitations of our model, the analytical expression of $R_0$ and the mathematically sound exploration of control strategies may make it relevant in the fields of epidemiology and public health. 
Our work highlights the potential causes of CCHF spread. 
The insights derived can pioneer the development of data-driven control measures modelling with scenarios and parameter values that are more realistic and adapted to a specific country or region. 
We expect that our work on CCHF spread and control measures may help to collect the necessary data related to CCHF and to further developing this and similar mathematical models and analyses.

\bibliography{mybibfile}

\end{document}